\documentclass[journal]{IEEEtran}
\usepackage{amsmath,amssymb,amsfonts}
\usepackage{algorithmic}
\usepackage{algorithm}
\usepackage{array}
\usepackage[caption=false]{subfig}
\usepackage{textcomp}
\usepackage{stfloats}
\usepackage{url}
\usepackage{verbatim}
\usepackage{graphicx}
\usepackage{cite}
\usepackage{acronym}
\usepackage{tikz}
\usepackage{multicol}
\usepackage{multirow}
\usepackage{mathrsfs}

\acrodef{cav}[CAV]{Connected Automated Vehicle}
\acrodef{gnss}[GNSS]{Global Navigation Satellite System}
\acrodef{v2v}[V2V]{Vehicle-To-Vehicle}
\acrodef{v2x}[V2X]{Vehicle-To-Everything}
\acrodef{dnn}[DNN]{Deep Neural Network}
\acrodef{cnn}[CNN]{Convolutional Neural Network}
\acrodef{iou}[IoU]{Intersection over Union}
\acrodef{nn}[NN]{Neural Network}
\acrodef{mec}[MEC]{Mobile Edge Cloud}
\acrodef{ml}[ML]{Machine Learning}
\acrodef{mmse}[MMSE]{Minimum Mean Square Error}
\acrodef{pdf}[pdf]{probability density function}
\acrodef{ap}[AP]{Average Precision}
\acrodef{cdf}[CDF]{cumulative distribution function}
\acrodef{cep}[CEP]{Circular Error Probable}
\acrodef{bev}[BEV]{Bird's Eye View}
\acrodef{kf}[KF]{Kalman Filter}
\acrodef{rmse}[RMSE]{Root Mean Square Error}
\acrodef{slam}[SLAM]{Simultaneous Localization and Mapping}
\acrodef{dl}[DL]{Deep Learning}
\acrodef{icp}[ICP]{Implicit Cooperative Positioning}
\acrodef{ssd}[SSD]{Single Shot Detector}
\acrodef{dsrc}[DSRC]{Dedicated Short-Range Communications}
\acrodef{uwb}[UWB]{Ultra Wide Band}
\acrodef{v2i}[V2I]{Vehicle-To-Infrastructure}
\acrodef{mpnn}[MPNN]{Message Passing Neural Network}
\acrodef{spa}[SPA]{Sum Product Algorithm}
\acrodef{mlp}[MLP]{Multi Layer Perceptron}
\acrodef{ekf}[EKF]{Extended Kalman Filter}
\acrodef{cls-mpnn}[CLS-MPNN]{Cooperative LiDAR Sensing with Message Passing Neural Network}
\acrodef{o-cls-mpnn}[O-CLS-MPNN]{Oracle CLS-MPNN}
\acrodef{icp-od-mpnn}[ICP-OD-MPNN]{Implicit Cooperative Positioning-Object Detection-Message Passing Neural Network}
\acrodef{o-icp-od-mpnn}[O-ICP-OD-MPNN]{Oracle ICP-OD-MPNN}
\acrodef{c-its}[C-ITS]{Cooperative Intelligent Transportation Systems}

\acrodef{nms}[NMS]{Non-Maximum Suppression}
\acrodef{ptpv2}[PTPv2]{Precision Time Protocol version 2}
\acrodef{gpu}[GPU]{Graphic Processing Unit}
\acrodef{rpn}[RPN]{Region Proposal Network}
\acrodef{fp}[FP]{False Positive}
\acrodef{da}[DA]{Data association}

\begin{document}
\bstctlcite{BSTcontrol}

\title{Deep Learning-based Cooperative LiDAR Sensing for Improved Vehicle Positioning}

\author{Luca Barbieri~\IEEEmembership{Member,~IEEE,} Bernardo Camajori Tedeschini~\IEEEmembership{Graduate Student Member,~IEEE,} \\
Mattia Brambilla~\IEEEmembership{Member,~IEEE,} Monica Nicoli~\IEEEmembership{Senior Member,~IEEE}
\thanks{This work is financed by the European Union—NextGenerationEU (National Sustainable Mobility Center CN00000023, Italian Ministry of University and Research Decree n. 1033—17/06/2022, Spokes 6 and 9). \\
L.\ Barbieri, B.\ Camajori Tedeschini and M.\ Brambilla are with Dipartimento di Elettronica, Informazione e Bioingegneria (DEIB), Politecnico di Milano, 20133 Milan, Italy \linebreak (e-mail:  luca1.barbieri@polimi.it, bernardo.camajori@polimi.it, mattia.brambilla@polimi.it).}
\thanks{M.\ Nicoli is with Dipartimento di Ingegneria Gestionale (DIG), Politecnico di Milano, 20133 Milan, Italy (e-mail:  monica.nicoli@polimi.it).}}



\maketitle

\begin{abstract}
Accurate positioning is known to be a fundamental requirement for the deployment of Connected Automated Vehicles (CAVs). To meet this need, a new emerging trend is represented by cooperative methods where vehicles fuse information coming from navigation and imaging sensors via Vehicle-to-Everything (V2X) communications for joint positioning and environmental perception. 
In line with this trend, this paper proposes a novel data-driven cooperative sensing framework, termed Cooperative LiDAR Sensing with Message Passing Neural Network (CLS-MPNN), where spatially-distributed vehicles collaborate in perceiving the environment via LiDAR sensors. Vehicles process their LiDAR point clouds using a Deep Neural Network (DNN), namely a 3D object detector, to identify and localize possible static objects present in the driving environment.
Data are then aggregated by a centralized infrastructure that performs Data Association (DA) using a Message Passing Neural Network (MPNN) and runs the Implicit Cooperative Positioning (ICP) algorithm. 
The proposed approach is evaluated using two realistic driving scenarios generated by a high-fidelity automated driving simulator. The results show that CLS-MPNN outperforms a conventional non-cooperative localization algorithm based on Global Navigation Satellite System (GNSS) and a state-of-the-art cooperative Simultaneous Localization and Mapping (SLAM) method while approaching the performances of an oracle system with ideal sensing and perfect association.

\end{abstract}


\section{Introduction}
\label{sec:introduction}



\IEEEPARstart{T}{he} demand for highly-accurate localization in \ac{c-its} has recently skyrocketed thanks to the advancements in the development of \ac{v2x} communication technologies~\cite{Grant:j11,V2X} which raised the attention to \acp{cav} and associated use cases~\cite{future_use_cases}.

Nowadays, most of the proposed solutions rely on the largely-available \ac{gnss}. 
However, even when augmented with inertial sensors, differential corrections, or multi-constellations receivers, \ac{gnss} is not able to satisfy the requirements of highly-accurate
positioning services~\cite{ETSITR122261,3GPP5GV2Xenhancedservices}, as it is heavily affected by poor visibility conditions that may occur in urban scenarios~\cite{gps_shortcomings}.
Novel design strategies (both from algorithmic and technological points of view) are being considered to improve vehicle positioning by integrating multiple onboard perception sensors~\cite{sensor_fusion,sensor_fusion2}, side information on the sensed environment~\cite{ConMazBar:J19} or exploiting \ac{v2x} communication technologies~\cite{5g_v2x}, rooted in 5G or beyond scenarios~\cite{6GV2X,ConMorLiuBar:J21,Del:J21, 10121016,tedeschini_latent_2023_2,italiano2023tutorial}.

Over the years, several positioning enhancements have been proposed to overcome the \ac{gnss} degradation in urban environments. 
\textcolor{black}{Among them, cooperative algorithms based on centralized or distributed frameworks~\cite{mattia_spa,mattia_icp_with_da,win_comm_magazine,patwari_signal_proc_magazine,win_proceedings,graph_diffusion} are gaining popularity as they improve vehicle positioning thanks to the sharing of location-dependent information over \ac{v2x} networks.}
Indeed, \ac{v2x} links allow aggregating the sensing information collected at multiple vehicles, which act as mobile probes of a larger distributed sensing system with an augmented view of the vehicular environment (compared to the ego vehicles) that facilitates the detection of objects along roads~\cite{survey_coop}.

\subsection{Related works}
\label{subsec:related_works}

Cooperative localization algorithms for vehicular applications have been proposed considering the exchange of \ac{gnss}~\cite{coop_loc_gps,coop_loc_gps2,8382292} or other types of radio measurements~\cite{ coop_loc_radio2,8950409,9145275,9405441,9417584,s21062048,coop_loc_radio}, and also with data fusion approaches~\cite{7822848,7996985,s20051413}.
In~\cite{coop_loc_gps}, a cooperative localization framework is proposed for correcting \ac{gnss} pseudoranges and improving the vehicle positioning accuracy, while~\cite{coop_loc_gps2} integrates \ac{gnss} and measurements obtained from \ac{dsrc} with a cubature Kalman Filter. 
Authors in~\cite{8382292} analyze the theoretical performances of inter-vehicle measurements extracted from \ac{gnss} observables. 
Besides \ac{gnss}-based solutions, radio measurements can also be used for enhancing localization accuracy.
Methods exploiting angle measurements are proposed in~\cite{coop_loc_radio2,8950409,9145275}, while~\cite{9405441,9417584} use the multipath information available from the channel impulse response. 
Other approaches, instead, focus on inter-distance measurements among vehicles~\cite{s21062048}, or on the \ac{gnss} carrier phase~\cite{coop_loc_radio}. 
\textcolor{black}{Lastly, data fusion methods where \ac{gnss} information and distances extracted from \ac{uwb} modules are combined together are studied in~\cite{7822848,7996985}, whereas~\cite{s20051413} combines distance and angle measurements.}

\textcolor{black}{Besides relying on either \ac{gnss}, \ac{v2x} radio measurements, or data fusion approaches, cooperative localization systems can also be implemented by exploiting imaging data coming from vehicle onboard perception sensors.}
A fast-developing technology that has been shown to provide promising detection and recognition capabilities for accurate environmental perception is LiDAR. 
The integration of LiDAR sensors within cooperative positioning schemes is therefore expected to bring significant advantages in terms of robustness and accuracy. 
Despite the LiDAR potentials, few cooperative approaches based on this technology have been studied~\cite{10215362, TedBraBarNic:C22,consistent_loc,9775023,coop_3d_mapping,edge_cooper,coop_obj_feature,coop_obj_feature_bw,coop_obj_infrastructure}.
\textcolor{black}{The works in~\cite{10215362,TedBraBarNic:C22} investigate the association issue across multiple vehicles, whereas~\cite{consistent_loc} develops a decentralized cooperative localization method to fuse LiDAR, \ac{gnss} and high-definition maps.
In~\cite{9775023}, a cooperative \ac{slam} approach is developed for vehicle pose estimation, while~\cite{coop_3d_mapping} proposes a cooperative mapping technique to fuse multiple LiDAR views into an improved unified map.
\textcolor{black}{Similarly, also~\cite{edge_cooper} develops a multi-vehicle cooperative scheme where vehicles share locally processed maps to improve cooperative perception.}
Other studies focus on collaborative \ac{ml} methods where intermediate \ac{nn} outputs are shared by vehicles~\cite{coop_obj_feature,coop_obj_feature_bw} and the road infrastructure~\cite{coop_obj_infrastructure}.
Despite the competitive performances provided by the aforementioned methods, they require the sharing of either raw or partially processed point clouds, which may not be feasible for bandwidth-limited \ac{v2x} applications, or they assume the availability of maps, limiting their application in rural or poorly visited areas.}

\textcolor{black}{In perception problems where passive, i.e., non-cooperative, targets are present, algorithms handling \ac{da} between measurements and targets are needed.
The pioneering Hungarian algorithm was proposed in \cite{hungarian_algorithm} for solving such problems.
It was then extended and superseded by approaches exploiting semantic information, which uniquely identifies the targets through visual features, such as color, size, or movement patterns~\cite{rakai_data_2022}.} 
In case measurements include only position-dependent parameters, such as distance or angle, the association problem becomes non-trivial to be solved. Solutions based on the \ac{spa} aim at efficiently computing the marginal association probabilities through the use of message passing over suitable factor graphs~\cite{kschischang_factor_2001,meyer_scalable_2020}, although they might suffer in case of cycle-graphs and non-Gaussian noise models. To improve upon these limitations, some works employ a double-loop algorithm, which forces the convergence of Gaussian belief propagation, and non-walksummable models~\cite{johnson_fixing_2009}, which involve the use of a preconditioner in an iterative method based on diagonal loading. Recent works exploit the capabilities of \ac{nn} over networks, also known as \ac{mpnn}~\cite{4700287,gilmer2017neural}, which learn the correct association directly from semantic data~\cite{TedBraBarNic:C22}. \acp{mpnn} can discern both linear and non-linear relations between input and output data, they are time scalable similar to \ac{spa}~\cite{zhou_graph_2020} and have demonstrated superior performance 
on loopy graphs, provided the availability of sufficient training data~\cite{10227084,yoon_inference_2019}.

\subsection{Contributions}

In this paper, we propose a novel data-driven cooperative localization method for \ac{gnss} augmentation in connected driving scenarios and suited for both rural and urban areas.
\textcolor{black}{The method builds on the \ac{icp} paradigm initially developed in~\cite{icp} and later extended in~\cite{mattia_icp_with_da} to address \ac{da}. 
Compared to those former works that assumed point-like detections obtained via generic sensing systems (e.g., LiDAR, camera, or radar), we here introduce a data-driven framework for accurate environmental perception from onboard LiDAR sensors at the vehicles.
Furthermore, we also propose to solve the \ac{da} problem via \acp{mpnn} to address the limitations of the \ac{spa}-based solution in~\cite{mattia_icp_with_da}.}
We refer to this proposed methodology as \ac{cls-mpnn}.

In the proposed \ac{cls-mpnn} solution, each vehicle is equipped with a \ac{dnn}, namely a 3D object detector, to efficiently process the LiDAR point clouds so as to recognize and localize passive static objects in the surroundings. 
\textcolor{black}{Note that the proposed framework can easily accommodate other types of sensors (e.g., cameras or radars) or multimodal solutions by properly modifying the 3D object detectors.}
Poles are selected as reference objects to be detected by vehicles as they are easily recognizable from the points clouds, widespread on most roads, and fixed.
\textcolor{black}{Nevertheless, the proposed framework is general enough to support other objects according to the scenario (e.g., rural or urban environments).}
The individual detections made by the vehicles are then exchanged with a centralized infrastructure, namely a \ac{mec}, via \ac{v2i} communications.
The \ac{mec} coherently combines the measurements provided by the vehicles by executing the \ac{mpnn}-based \ac{da} algorithm and then it runs the \ac{icp} method for refining both objects and vehicle positions.
\textcolor{black}{We consider a \ac{mec}-centralized processing architecture as it provides the highest performances and it can be implemented at the road infrastructure, even if it has the drawback of being the critical single point of failure of the system.
The method is general enough to be adapted to distributed architectures, following e.g., the approach in~\cite{mattia_icp_with_da}.} 



\textcolor{black}{With respect to previous works that either rely on partially processed \cite{coop_obj_feature,coop_obj_feature_bw,coop_obj_infrastructure} or raw LiDAR sharing \cite{9775023,coop_3d_mapping}, the proposed method foresees only the exchange of the detected objects and vehicles positions, thereby substantially limiting the communication overhead.
Furthermore, the developed method does not rely on maps of the driving environment, as in \cite{consistent_loc}, making the cooperative approach potentially applicable to any driving situation, especially when high-definition maps are unavailable.}
\textcolor{black}{The proposed solution improves upon all our previous works, which provided an initial proof of concept using only single-stage detectors~\cite{Barbieri2023_ICC,BarTedBraNic:C23}, or focused on studying the \ac{da} problem alone~\cite{TedBraBarNic:C22,10215362}. 
Herein, we provide a more complete and robust solution that jointly addresses the \ac{da} problem and improves the localization accuracy thanks to the cooperation among vehicles.
Specifically, the main contributions are as follows: 
}
\begin{itemize}
    \item \textcolor{black}{We embed the \ac{da} in a cooperative mechanism by extending the method in~\cite{TedBraBarNic:C22,10215362} to enable tracking over time.} 
    \item \textcolor{black}{We extend our frameworks in~\cite{10215362,TedBraBarNic:C22,Barbieri2023_ICC,BarTedBraNic:C23} by integrating both single-stage and double-stage 3D object detectors to provide a comprehensive cooperative localization platform.} 
    \item \textcolor{black}{We introduce a comparison with state-of-the-art cooperative \ac{slam} approaches to show the benefits of the proposal in terms of positioning accuracy as well as in reducing the communication overhead.} 
    \item \textcolor{black}{We study the generalization abilities of \ac{cls-mpnn} in new driving scenarios where neither the 3D object detector nor the \ac{mpnn} were trained on. Besides, we also characterize the ability of \ac{cls-mpnn} to work with cheaper, lower-resolution LiDAR sensors and its robustness against false alarms.} 
\end{itemize}

The overall goal is to provide the achievable performance of a cooperative system that uses LiDAR sensing to detect static objects in the road environment for vehicle positioning augmentation.
\textcolor{black}{For this reason, we assume noisy detections neglecting false alarms or \acp{fp} (i.e., undesired objects that are either erroneously detected by the \ac{dnn} model or generate backscattering with high intensity).}
\textcolor{black}{Experimental results show that \ac{cls-mpnn} substantially outperforms an ego-vehicle positioning method based on \ac{gnss} tracking for different traffic densities and LiDAR resolutions while approaching the performances of an ideal cooperative LiDAR sensing system with perfect \ac{da} that serves as a possible lower bound on the performances.}
\textcolor{black}{Additionally, the proposal is shown to be advantageous even when compared with state-of-the-art cooperative \ac{slam} algorithms and when the \acp{fp} are not removed.} 
Lastly, when testing is performed in a scenario different from the one used for training, the cooperative approach still largely outperforms the ego \ac{gnss} tracking one, even though more missed detections are observed. 

\textcolor{black}{The paper is organized as follows. Sec.~\ref{sec:system_model} details the system model considered throughout the paper.
Sec.~\ref{sec:dnn_sensing} presents the \ac{ml} methods adopted for environmental perception, while Sec.~\ref{sec:coop_lidar_pos} describes the Bayesian tracking framework and the \ac{mpnn}-based \ac{da}. 
Sec.~\ref{sec:challenges} discusses potential implementation challenges.
Sec.~\ref{sec:nn_results} highlights the performances of the 3D object detectors and the \ac{mpnn}, while Sec.~\ref{sec:coop_loc_results} presents the result characterizing the proposed cooperative localization framework. 
Lastly, Sec.~\ref{sec:conclusions} draws conclusions.} 

\begin{figure}
    \centering
    \subfloat[\label{fig:scenario_carla}]{
        \includegraphics[width=0.95\columnwidth]{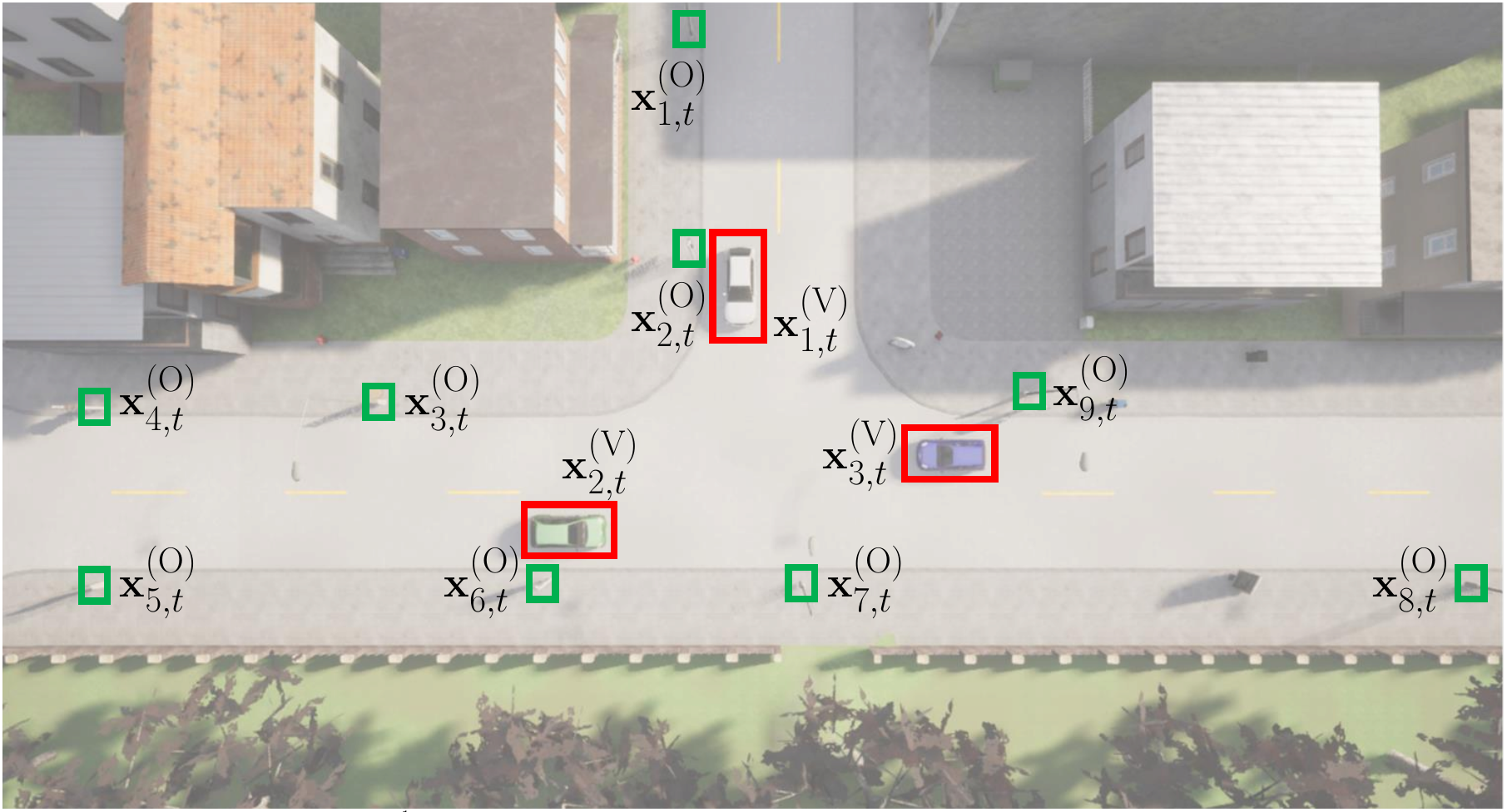}
    }

    \subfloat[\label{fig:scenario_point_clouds}]{
    \includegraphics[width=0.95\columnwidth]{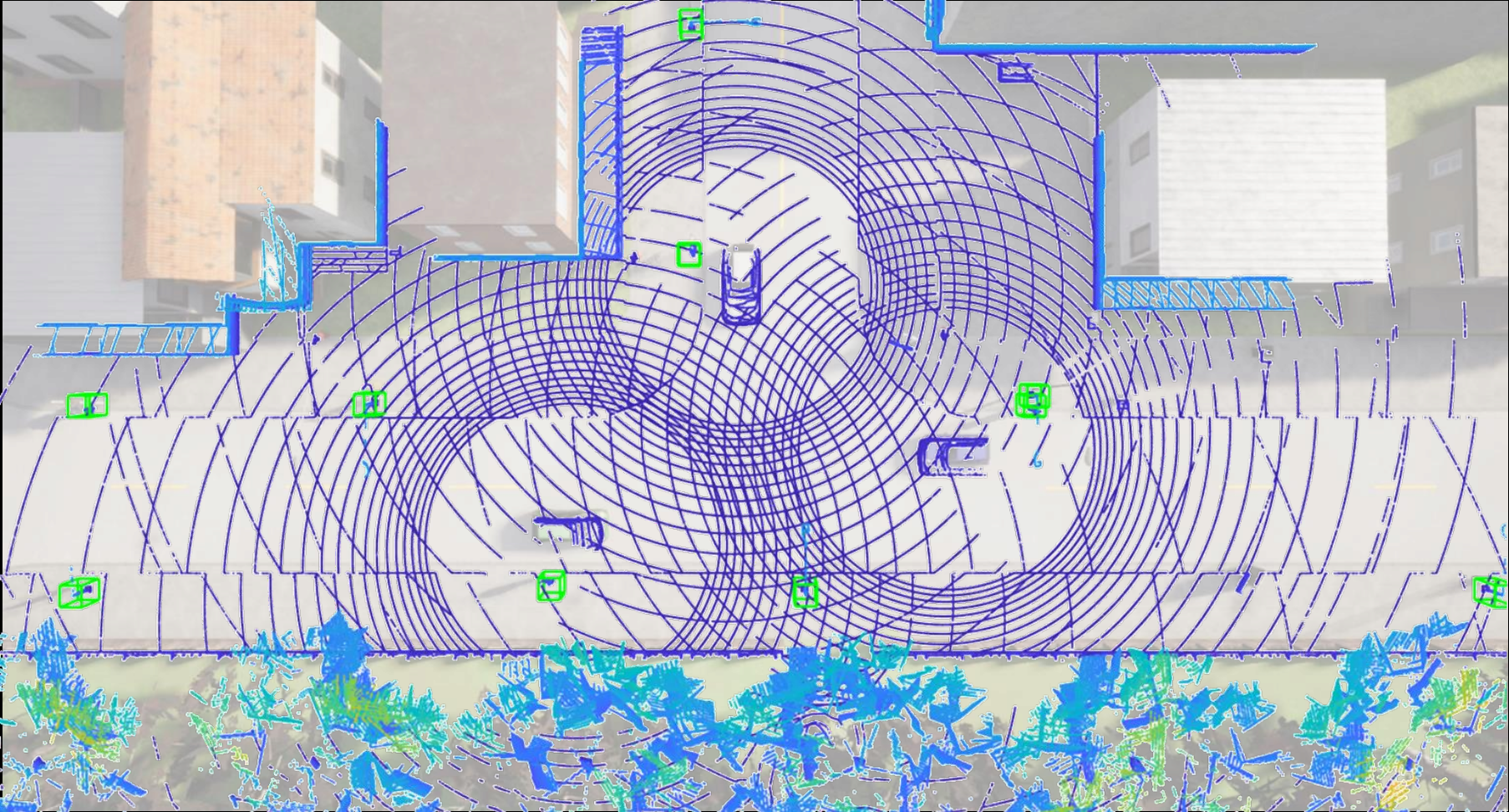}
    }
\vspace{-2mm}    \caption{\textcolor{black}{Representation of vehicular sensing scenario with $N_v=3$ vehicles and $N_o=9$ non-cooperative objects (poles). (a) Top view of a scenario extracted from CARLA. (b) LiDAR point clouds captured by the vehicles.}}
    \label{fig:enter-label}
    \vspace{-4mm}
\end{figure}

\section{Vehicular scenario and models}
\label{sec:system_model}

In this section, we introduce the connected vehicular scenario and the main assumptions used in the paper for the design of the cooperative localization solution. 
More specifically, Sec.~\ref{subsec:loc_model} defines the models for the vehicle dynamics and the localization measurements available for cooperative positioning refinement; whereas Sec.~\ref{subsec:data_association_model} presents the modeling of the \ac{da} problem and the related variables. 

\subsection{Localization model}
\label{subsec:loc_model}

\textcolor{black}{The road scenario comprises a set $\mathcal{V} = \{1, \ldots, N_v\}$ of $N_v$ vehicles moving over a 3D space.}
At discrete time $t$, each vehicle $v \in \mathcal{V}$ is characterized by a 3D position $\mathbf{u}_{v,t}^{(\text{V})} = [u_{v,x,t}^{(\text{V})}\, u_{v,y,t}^{(\text{V})} \, u_{v,z,t}^{(\text{V})}]^{\mathrm{T}}$ and a 3D velocity $\mathbf{v}_{v,t}^{(\text{V})} = [v_{v,x,t}^{(\text{V})}\, v_{v,y,t}^{(\text{V})} \, 0]^{\mathrm{T}}$, which are collected into the time-varying state $\mathbf{x}_{v,t}^{(\text{V})} = [\mathbf{u}_{v,t}^{{(\text{V})}^{\mathrm{T}}}\, \mathbf{v}_{v,t}^{{(\text{V})}^{\mathrm{T}}} ]^{\mathrm{T}}$. 
Additionally, each vehicle can establish a \ac{v2i} communication link with a centralized infrastructure, namely a \ac{mec}, for exchanging positioning and sensing information. 
\textcolor{black}{The vehicle state is assumed to evolve according to the following first-order Markov model, namely the constant velocity~\cite{motion_models}} 
\begin{equation}
\mathbf{x}_{v,t}^{(\text{V})} = \mathbf{F} \mathbf{x}_{v,t-1}^{(\text{V})} + \mathbf{K}\mathbf{w}_{v,t-1}^{(\text{V})} \, ,
\label{eq:motion_model_vehicle}
\end{equation}
with $\mathbf{F}\!\!=\!\! [\mathbf{I}_3, \! T_s \mathbf{I}_3; \! \mathbf{0}_{3\times3}, \! \mathbf{I}_3]$ 
and $\mathbf{K}\!\!=\!\![0.5 T_s^2 \mathbf{I}_2;\! \mathbf{0}_{1 \times 2};\!  T_s \mathbf{I}_2; \!\mathbf{0}_{1 \times 2}]$, while $\mathbf{w}_{v,t-1}^{(\text{V})}$ is a zero-mean Gaussian driving process modeling the acceleration uncertainty with covariance $\mathbf{Q}^{(\text{V})}_{v,t-1} = \sigma_{Q}^2 \mathbf{I}_2$ and $T_s$ the sampling interval.
Besides vehicles, the scenario contains a set $\mathcal{O} = \{1, \ldots, N_o\}$ of $N_o$ non-cooperative static objects that can be detected by nearby vehicles.
Ideally, ad-hoc objects should be placed in the driving environment to facilitate their detection from the imaging sensors of the vehicles. 
For this initial study, we select poles as detectable objects since they are extremely common, easily recognizable and fixed.
\textcolor{black}{Each pole is described by a position state 
$\mathbf{x}_{k,t}^{(\text{O})} =
[x_{k,x,t}^{(\text{O})}   \, x_{k,y,t}^{(\text{O})} \, x_{k,z,t}^{(\text{O})}]^{\mathrm{T}}$ here assumed to be time-invariant:
$\mathbf{x}_{k,t}^{(\text{O})} = \mathbf{x}_{k,t-1}^{(\text{O})}$.}
\textcolor{black}{A visualization of a portion of the scenario from the software used for simulation, i.e., CARLA~\cite{pmlr-v78-dosovitskiy17a}, is provided in Fig.~\ref{fig:enter-label}.}

The vehicles are able to estimate their own position using navigation units, e.g., \ac{gnss}, which are assumed to be always available.
The \ac{gnss} measurement at vehicle $v$ is modeled as 
\begin{equation}
\boldsymbol{\rho}_{v,t}^{(\text{V})} = \mathbf{T}\mathbf{x}_{v,t}^{(\text{V})} + \mathbf{n}_{v,t}^{(\text{V})} \, ,
\end{equation}
where $\mathbf{T} = [\mathbf{I}_{2}\,  \mathbf{0}_{2 \times 4}]$, while $\mathbf{n}_{v.t}^{(\text{V})}$ is a zero-mean Gaussian measurement error with covariance $\mathbf{R}_{v,t}^{(\text{V})} = \sigma_{\text{P}}^{(\text{V})^2}\mathbf{I}_2$ with $\sigma_{\text{P}}^{(\text{V})}$ denoting the \ac{gnss} position accuracy. 

Each vehicle is also able to detect possible objects present in the surroundings via LiDAR sensors.
Specifically, the point cloud $\mathbf{P}_{v,t}$ captured at time $t$ by vehicle $v$ is processed by a 3D object detector, whose task is to generate a candidate set of bounding boxes physically encapsulating the detected poles.
Such set is denoted as $\mathcal{O}_{v,t} = \{ k \in \mathcal{O}: \|\mathbf{u}_{v,t}^{(\text{V})} - \mathbf{x}_{k,t}^{(\text{O})} \| \leq R_s\} \subseteq \mathcal{O}$ with cardinality $|\mathcal{O}_{v,t}|$,  where $R_s$ is the LiDAR sensing range (here assumed to be equal for all vehicles). 

The 3D object detector provides two different types of measurements.
The first one refers to the 3D position of the bounding boxes' centroids and is given by 
\begin{align}
\boldsymbol{\rho}_{v,t}^{\text{(O)}} = \{\boldsymbol{\rho}_{v,k,t}^{\text{(O)}} \}_{k = 1}^{|\mathcal{O}_{v,t}|} , 
\label{eq:meas_centriods}
\end{align}
where $\boldsymbol{\rho}_{v,k,t}^{\text{(O)}}$, with $k = 1, \ldots, |\mathcal{O}_{v,t}|$, denotes the distance between the $k$-th bounding box center and the $v$-th vehicle position at time $t$. 
The second one, instead, represents the detected poles as a collection of eight corners
\begin{equation}
\mathbf{z}_{v,t}^{\text{(O)}} = \left\{\mathbf{z}_{v,k,t}^{\text{(O)}} \right\}_{k = 1}^{|\mathcal{O}_{v,t}|} = \left\{\left[\mathbf{z}_{v,k,c,t}^{\text{(O)}}  \right]_{c=1}^{8} \right\}_{k = 1}^{|\mathcal{O}_{v,t}|} \, ,
\label{eq:meas_corners}
\end{equation}
where $\mathbf{z}_{v,k,c,t}^{\text{(O)}}$ represents the 3D coordinate of a corner point.
\textcolor{black}{We specifically represent objects as bounding boxes as they can be easily obtained from the output provided by the 3D object detector (as highlighted in Sec. \ref{sec:dnn_sensing}).
Note that other choices can still be used by converting the output of the 3D object detectors in an appropriate format.}
For \ac{da} purposes, we also define the sets of estimated absolute locations of the centroids and the  bounding box corners as
\begin{align}
\overline{\boldsymbol{\rho}}_{v,t}^{\text{(O)}} &= \left\{\overline{\boldsymbol{\rho}}_{v,k,t}^{\text{(O)}}\right\}_{k = 1}^{|\mathcal{O}_{v,t}|} = \left\{\boldsymbol{\rho}_{v,k,t}^{\text{(O)}} + \widehat{\mathbf{u}}_{v,t}^{(\text{V})} \right\}_{k = 1}^{|\mathcal{O}_{v,t}|} \, ,
\label{eq:setOfCenters}
    \\
\overline{\mathbf{z}}_{v,t}^{\text{(O)}} &= \left\{\overline{\mathbf{z}}_{v,k,t}^{\text{(O)}} \right\}_{k=1}^{|\mathcal{O}_{v,t}|} = \left\{\left[\mathbf{z}_{v,k,c,t}^{\text{(O)}}\right]_{c=1}^{8} + \widehat{\mathbf{u}}_{v,t}^{(\text{V})} \right\}_{k = 1}^{|\mathcal{O}_{v,t}|} \, , 
\label{eq:setOfCorners}    
\end{align}
where $\widehat{\mathbf{u}}_{v,t}^{(\text{V})}$ is the predicted position of vehicle $v$ at time~$t$ from its previous state estimate $\widehat{\mathbf{x}}_{v,t-1}^{(\text{V})}$ according to the motion model in \eqref{eq:motion_model_vehicle}.
The definition of the sets \eqref{eq:setOfCenters}-\eqref{eq:setOfCorners} is required to enable the coherent fusion of the detected objects across vehicles by performing the \ac{da} step in a common reference system.
Specifically, set  $\overline{\mathbf{z}}_{v,t}^{\text{(O)}}$ is useful for solving the \ac{da} problem as it represents the bounding boxes accounting for both orientation and dimension. 
By contrast, set $\overline{\boldsymbol{\rho}}_{v,t}^{\text{(O)}}$ enables to estimate the absolute position $\widehat{\mathbf{x}}_{o,t}^{(\text{O})}$ of each object $ o \in \mathcal{O}_{v,t}$.

\subsection{Data association model}
\label{subsec:data_association_model}

A fundamental step required to coherently aggregate the bounding box measurements generated by each vehicle is the \ac{da}, which enables the identification of bounding boxes relating to the same object across multiple vehicles.
\textcolor{black}{This operation is typically affected by several errors.
Examples include the intrinsic error on the point clouds (in the order of centimeters), the error introduced by the object detection process (for some reference values please see the analysis provided in Sec. \ref{sec:nn_results}), and the accuracy of the onboard positioning system (e.g., \ac{gnss}), which represents the highest source of error (in the order of meters).
}
Overall, the combination of these noise sources makes solving the \ac{da} problem non-trivial. 
To address these shortcomings, we propose to employ \acp{mpnn}, as detailed hereafter. 

Given the set $\mathcal{Z}_t = \bigcup_{v=1}^{N_v} \overline{\mathbf{z}}_{v,t}^{(\text{O})}$ comprising all the bounding box corners generated by all vehicles at time $t$, the \ac{mpnn} input is modeled starting from $\mathcal{Z}_t$ so as to obtain a directed graph $\mathcal{G}=(\mathcal{N}, \mathcal{E})$, where $\mathcal{N}$ and $\mathcal{E}$ are the sets of nodes and edges, respectively. 
The graph $\mathcal{G}$ is obtained via a mapping function $\Phi_t : \mathcal{N} \rightarrow \mathcal{Z}_t \times \mathcal{V}$ such that each bounding box $k = 1 , \dots , |\mathcal{O}_{v,t}| $
of vehicle $v$ is associated to a graph node $n \in \mathcal{N}$, while the edge $(n,m) \in \mathcal{E}$, with $n\neq m$, denotes a potential association between the bounding boxes encoded by nodes $n$ and $m$. 
Additionally, we introduce the association variable $\mathtt{a}_{n \to m} \in \{0,1\}$, where $\mathtt{a}_{n \to m}=1$ denotes the existence of the edge $(n,m)$, or equivalently that the bounding boxes associated with nodes $n$ and $m$ refer to the same object, while $\mathtt{a}_{n \to m}=0$ indicates that no edge exists between nodes $n$ and $m$. 
The goal of the MPNN is to infer the association variables $\widehat{\mathtt{a}}_{n \to m} \in \{0,1\}$ by considering all the potential bounding box combinations for all vehicles. 
To avoid associating two (or more) bounding boxes estimated by the same vehicle, we impose $\widehat{\mathtt{a}}_{n \to m} = 0 $ whenever the  mappings $\Phi_t(n)$ and $\Phi_t(m)$ refer to the same vehicle.

Besides the measurement-to-measurement association variable $\mathtt{a}_{n \to m} $, we also define the binary vector $\boldsymbol{\alpha}_t = [\alpha_{v,o,t}]_{v \in \mathcal{V}, o \in \mathcal{O}_{t}}$  collecting all the measurement-to-object association variables of all vehicles at time~$t$, where $\mathcal{O}_t = \bigcup_{v=1}^{N_v} \mathcal{O}_{v,t} $ is the set of indexes of all detected objects at time $t$.
Given the output of the \ac{mpnn}, another step is required to pair the (associated) measurements with the objects and obtain an estimate of the measurement-to-object association variable $\widehat{\boldsymbol\alpha}_t$. 
This step will be detailed in Sec.~\ref{subsec:bayesian_tracking}.

\section{Deep learning-based lidar sensing}
\label{sec:dnn_sensing}

In this section, we detail the strategies used for enabling each vehicle to perceive the driving environment. 
In particular, \ac{dl}-based 3D object detectors are used to efficiently process the LiDAR point clouds and obtain the position of the static objects in the vehicles' surroundings. 
\textcolor{black}{Given the ego-vehicle LiDAR data, the 3D object detectors provide at the output a set $\mathscr{B}= \{1, \ldots, N_{B}\}$ of $N_B$ candidate bounding boxes physically encapsulating the detected objects.
Each bounding box is compactly described as $\widehat{\mathbf{b}}_{j}= [\widehat{\mathbf{u}}_{j}^{(\text{B})}\, \widehat{\mathbf{d}}_{j}\,  \widehat{\gamma}_{j} ]^{\mathrm{T}}$ with $j = 1, \dots , N_{B}$, where $\widehat{\mathbf{u}}_{j}^{(\text{B})}~=~ [\widehat{u}_{x,j}^{(\text{B})}\, \widehat{u}_{y,j}^{(\text{B})}\, \widehat{u}_{z,j}^{(\text{B})}]$ is the estimated object's centroid, $\widehat{\mathbf{d}}_{j} = [\widehat{\ell}_{j} \, \widehat{w}_{j} \, \widehat{h}_{j}]$ contains the object's dimensions, namely the length, width and height, while $\widehat{\gamma}_{j}$ is the yaw angle.
The outputs provided by the detectors can then be suitably converted to obtain the measurements defined in \eqref{eq:meas_centriods}-\eqref{eq:meas_corners}.  
Besides, each bounding box is associated with a score probability, ranging from 0 up to 1, that indicates the (soft) confidence of the detector.
Herein, we assume that all bounding boxes provided by the 3D object detectors refer only to poles present in the driving environment, namely all \acp{fp} are filtered out.
Nevertheless, the score probability can be used to suppress the \acp{fp} as analyzed in Sec. \ref{subsec:town02_results}.}


To design the cooperative localization framework, we employ two state-of-the-art 3D object detection methods with increasing complexity, namely PointPillars and Part-$\text{A}^2$ that are based on the single-stage and double-state paradigms, respectively \cite{dl_point_clouds}.
In the following, we initially describe the inner workings of PointPillars (Sec.~\ref{subsec:pointpillars}) and then present the Part-$\text{A}^2$ method (Sec.~\ref{subsec:parta2}).

\subsection{PointPillars}
\label{subsec:pointpillars}

PointPillars is a single-stage object detector that processes the input point cloud and extracts important features organized in pillars through the employment of PointNet \cite{pointnet} modules.
The encoded features are then processed by a standard \ac{rpn} to obtain the final predictions.
In the following, we explain its inner working principles.

\textcolor{black}{Starting from a single-vehicle point cloud $\mathbf{P} = [\mathbf{p}_{1} \cdots \mathbf{p}_{N_{\ell p}}]^{\mathrm{T}}
\triangleq \mathbf{P}_{v,t}$\footnote{
\textcolor{black}{We remove the dependency on the vehicle and time indexes to simplify the disclosure in this section.}}of $N_{\ell p}$ points, PointPillars converts it into an image-based representation.
Specifically, $\mathbf{P}$ is discretized into an evenly-spaced $x$-$y$ grid of $\mathcal{P}$ non-empty pillars, each containing $N_p$ points\footnote{\textcolor{black}{This is done by padding pillars with zeros until they have exactly $N_p$ points or randomly removing points from pillars having more than $N_p$ points.}}.
The output is then fed to a simplified PointNet architecture as presented in \cite{point_pillars} to obtain a 2D pseudo-image of size $C \times H \times W$, with $C$, $H$ and $W$ denoting the number of channels, height and width of the canvas, respectively.
The pseudo-image is used by a \ac{rpn} to estimate the 3D bounding boxes. 
Specifically, the \ac{rpn} is composed by a backbone, whose task is to compute important features, and a detection head that uses the backbone's features to generate the final predictions. 
In PointPillars,  the \ac{ssd} head with pre-defined anchor (or prior) boxes is used.  
The matching strategy \cite{voxel_net} used by PointPillars assigns a  positive match to the predicted bounding box having the highest 2D \ac{iou} with the ground-truth box or a $\text{2D} \,  \text{IoU} > 0.5$. On the other hand, predicted boxes with $\text{2D}\, \text{IoU} < 0.35$ are deemed as negative samples. All other anchors are ignored for the loss computation. 
During inference, an axis-aligned \ac{nms} stage~\cite{voxel_net} with an \ac{iou} threshold of 0.5 is applied to filter out overlapping boxes that refer to the same object.
Finally, the \ac{nn} parameters of PointPillars are updated according to the losses presented in \cite{point_pillars} using the same configuration parameters.}

\subsection{Part-$A^{2}$}
\label{subsec:parta2}

Part-$A^{2}$ is a double-stage 3D object detector that introduces the concept of intra-object part estimation and aggregation.
It extends PointRCNN~\cite{point_rcnn} to concurrently learn segmentation masks for discerning foreground points, classify their position inside the bounding box and generate 3D proposals, i.e., rough bounding box estimates.
The proposals are then fed into a part-aggregation network that generates the final predictions, exploiting also the intra-object part information produced in the preceding stage. 
Note that we use in our experiments the Part-$\text{A}^2$-Anchor, which uses prior boxes for the proposal generation process. 
In the following, we briefly describe the overall architecture and how the bounding boxes are estimated.

\textcolor{black}{In contrast to PointPillars, Part-$\text{A}^2$ process the single-vehicle point cloud $\mathbf{P}$ using voxelization. 
In particular, a regular 3D grid is computed starting from $\mathbf{P}$, and the points associated with each grid point are grouped and averaged, leading to a new point cloud representation.
Then, a UNet-based encoder-decoder~\cite{unet} segments the foreground points, i.e., points belonging to the ground truth bounding boxes, from all the other points.
Concurrently, an additional module is responsible for learning the intra-object part information.
Then, a \ac{rpn} is appended for generating the 3D proposals (with pre-defined prior boxes).
The matching strategy uses bounding boxes with $\text{3D} \, \text{IoU} > 0.5$ as positive anchors, while boxes having $\text{3D} \, \text{IoU} < 0.35$ are deemed as negative.
All other proposals are neglected during loss computation.
After this step, only $64$ positive and $64$ negative anchor boxes are retained for further refinement.}

\textcolor{black}{Once the proposals have been generated, a second network is responsible for improving the detection accuracy.
This module takes as input the intra-object part locations estimated before, together with the proposals and the associated learned features, and refines the estimated bounding boxes. 
At first, the points falling within each proposal are transformed into a canonical coordinate system where the $x$ and $y$ axes are approximately parallel to the ground plane, while the $z$ axis is the same as the input point cloud. 
This reduces the impact of different rotation angles and locations of the proposals, leading to a better bounding box regression.
Then, a Region of Interest (RoI)-aware pooling strategy is used to aggregate the features produced during the previous stage as in \cite{parta2}.
Finally, the pooled features are fed to sparse convolutional layers whose outputs are used by two separate branches for bounding box regression and classification.
The confidence scores are assigned by a 3D IoU guided scoring strategy~\cite{parta2}.
A confidence of 0 is assigned to boxes having $\text{3D} \,\text{IoU} < 0.25$, while for boxes with $\text{3D} \,\text{IoU} > 0.75$ the confidence is set to 1.
In all other cases, the confidence is computed as $2\text{IoU} - 0.25$.
During inference, a rotational \ac{nms} strategy with \ac{iou} threshold of 0.7 is applied to filter out overlapping boxes that refer to the same object. 
Part-$\text{A}^2$ is trained in an end-to-end manner by exploiting the losses presented in \cite{parta2}}.

Both \ac{dnn} detectors are trained in a centralized manner and then deployed at the vehicles. In this way, each vehicle $v$ can estimate the bounding boxes from the point cloud  $\mathbf{P}_{v,t}$ and obtain the corresponding sets $\boldsymbol{\rho}_{v,t}^{\text{(O)}}$ and $\mathbf{z}_{v,t}^{\text{(O)}}$.

\section{Cooperative lidar positioning}
\label{sec:coop_lidar_pos}


\begin{figure}
    \centering
    
    \begin{tikzpicture}
        \node[]at(0,0.1){\includegraphics[width=\linewidth]{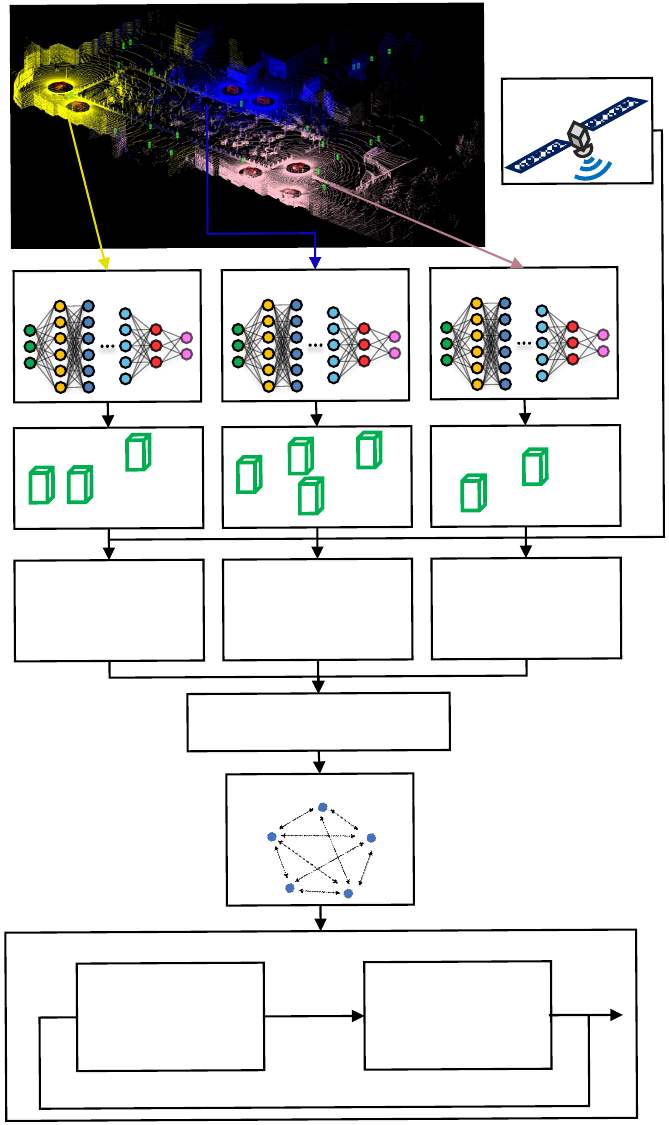}};
        \node[align=center]at(-1.4,7.65){LiDAR point cloud};
        \node[align=center]at(3.2,6.35){GNSS};
        \node[align=center]at(2.6,3.8){DNN model};
        \node[align=center]at(-0.2,3.8){DNN model};
        \node[align=center]at(-3,3.8){DNN model};
        \node[align=center]at(2.6,-0.5){\shortstack{measurement \\ aggregation \\ at vehicle}};
        \node[align=center]at(-0.2,-0.5){\shortstack{measurement \\ aggregation \\ at vehicle}};
        \node[align=center]at(-3,-0.5){\shortstack{measurement \\ aggregation \\ at vehicle}};
        \node[align=center]at(-0.2,-2){MEC data aggregation};
        \node[align=center]at(-0.2,-2.9){Data association};
       \node[align=center]at(-1.7,-4.99){Cooperative Bayesian localization};
        \node[align=center]at(-2.2,-5.85){\shortstack{Vehicle/object \\ state estimate}};
        \node[align=center]at(1.65,-5.85){\shortstack{Vehicle/object \\ state prediction}};

    \end{tikzpicture}
    \vspace{-8mm}
    \caption{\textcolor{black}{Block scheme of the proposed CLS-MPNN localization framework. Each vehicle processes its local LiDAR point cloud via 3D object detectors to localize passive objects. The GNSS and the bounding box measurements  are aggregated and shared with the \ac{mec}, which performs \ac{da} and cooperatively tracks both objects and vehicle states.}}
    \label{fig:icp_scheme}
\end{figure}
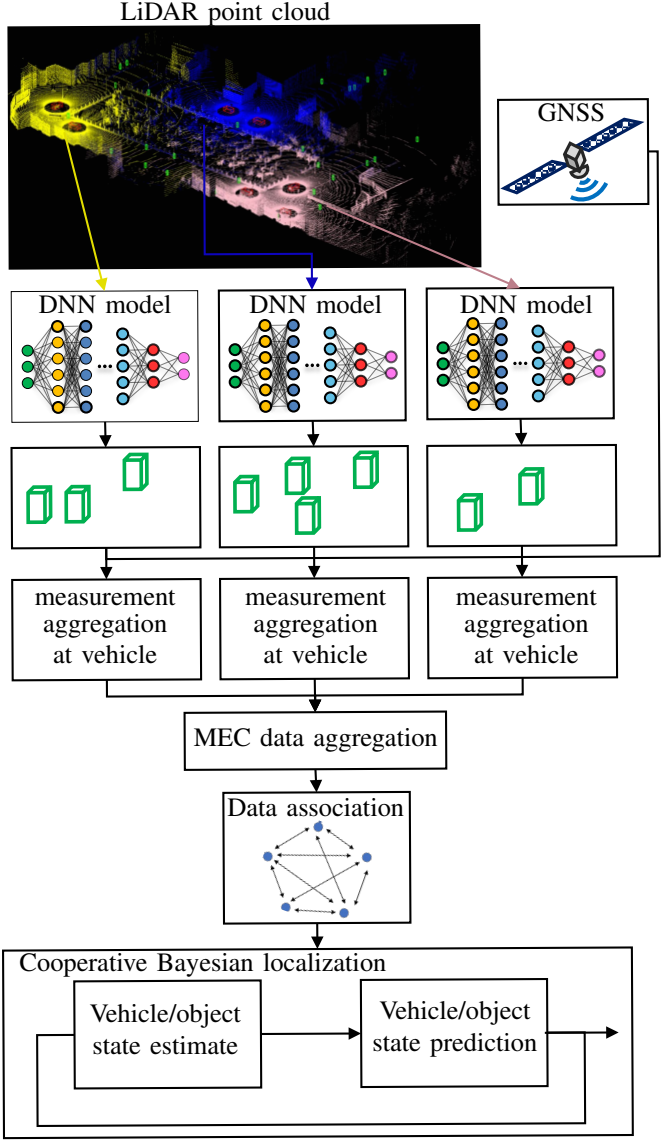

In this section, we present the \ac{cls-mpnn} localization framework employed for cooperatively localizing both objects and vehicles and tracking their states across time. 
Sec.~\ref{subsec:bayesian_tracking} introduces the Bayesian tracking solution, then we describe how the \ac{da} problem is addressed.
In particular, we solve the \ac{da} problem in two steps: the first one enables the estimation of the measurement-to-measurement association variable via the \ac{mpnn} (Sec.~\ref{subsec:meas_to_meas_association}), while the second one estimates the measurement-to-target associations (Sec.~\ref{subsec:meas_to_target_association}).
The overall  block scheme of the \ac{cls-mpnn} positioning framework is summarized in Fig.~\ref{fig:icp_scheme}. 


\subsection{Bayesian tracking solution}
\label{subsec:bayesian_tracking}

\textcolor{black}{We assume that at each time $t$ all vehicles $v \in \mathcal{V}$ are able to send to the road infrastructure via \ac{v2i} links their position measurements $\boldsymbol{\rho}_{v,t}^{\text{(V)}}$ together with the sets $\mathbf{z}_{v,t}^{\text{(O)}}$ and $\boldsymbol{\rho}_{v,t}^{\text{(O)}}$ of detected poles}. 
The \ac{mec} is then in charge of cooperatively
tracking the state evolution of the vehicles and the objects over
consecutive time instants.
\textcolor{black}{To do so, we resort to a Bayesian approach where we are interested in computing the joint posterior \ac{pdf} of vehicles and objects, namely $p(\boldsymbol{\theta}_{t}| \boldsymbol{\rho}_{1:t}^{(\text{V})}, \boldsymbol{\rho}_{1:t}^{(\text{O})} )$, where $\boldsymbol\theta_t = [\mathbf{x}_{t}^{(\text{V})^{\text{T}}} \, \mathbf{x}_{t}^{(\text{O})^{\text{T}}}]^{\text{T}}$ collects all the vehicles $\mathbf{x}_{t}^{(\text{V})}$ and object $\mathbf{x}_{t}^{(\text{O})}$ states, while $\boldsymbol{\rho}_{1:t}^{(\text{V})}$ and $\boldsymbol{\rho}_{1:t}^{\text{(O)}}$ aggregate all the measurements
$\boldsymbol{\rho}_{v,t}^{(\text{V})}$ and
$\boldsymbol{\rho}_{v,t}^{\text{(O)}}$, respectively, of all vehicles up to time~$t$.
The overall state can be then estimated using the \ac{mmse} criterion, leading to
\begin{equation}
    \widehat{\boldsymbol{\theta}}_t = \int \boldsymbol{\theta}_t \,p(\boldsymbol{\theta}_{t}| \boldsymbol{\rho}_{1:t}^{(\text{V})}, \boldsymbol{\rho}_{1:t}^{(\text{O})} )
    \,
    \mathrm{d}\boldsymbol{\theta}_t \,.
\end{equation}
Such estimation process entails computing  $p(\boldsymbol{\theta}_{t}| \boldsymbol{\rho}_{1:t}^{(\text{V})}, \boldsymbol{\rho}_{1:t}^{(\text{O})} )$, which needs to be evaluated after the \ac{da} is resolved. 
By conditioning with respect to the association variable $\boldsymbol{\alpha}_t$, the posterior is rewritten as 
\begin{align}
p(\boldsymbol{\theta}_t | \boldsymbol{\rho}_{1:t}^{(\text{V})} , \boldsymbol{\rho}_{1:t}^{\text{(O)}} )
= 
\sum_{\boldsymbol{\alpha}_t}
p(\boldsymbol{\theta}_t | \boldsymbol{\alpha}_{t}, \boldsymbol{\rho}_{1:t}^{(\text{V})} , \boldsymbol{\rho}_{1:t}^{\text{(O)}} )
\,
p(\boldsymbol{\alpha}_t | \boldsymbol{\rho}_{1:t}^{(\text{V})} , \boldsymbol{\rho}_{1:t}^{\text{(O)}} ) \,.
\end{align}
where $p(\boldsymbol{\theta}_t | \boldsymbol{\alpha}_{t}, \boldsymbol{\rho}_{1:t}^{(\text{V})} , \boldsymbol{\rho}_{1:t}^{\text{(O)}} )$ is evaluated according to the mathematical formulation  in \cite{mattia_icp_with_da}, while $p(\boldsymbol{\alpha}_t | \boldsymbol{\rho}_{1:t}^{(\text{V})} , \boldsymbol{\rho}_{1:t}^{\text{(O)}} )$ is obtained by solving the \ac{da}. 
Specifically, we  split the computation of
$p(\boldsymbol{\alpha}_t | \boldsymbol{\rho}_{1:t}^{(\text{V})} , \boldsymbol{\rho}_{1:t}^{\text{(O)}} )$ into two steps. 
The first one uses the \ac{mpnn} to pair bounding boxes hypothesized to refer to the same object while the second evaluates the measurement-to-target association, providing as output the pdf  $p(\boldsymbol{\alpha}_t | \boldsymbol{\rho}_{1:t}^{(\text{V})} , \boldsymbol{\rho}_{1:t}^{\text{(O)}} )$,
}

\subsection{Measurement-to-measurement association}
\label{subsec:meas_to_meas_association}

\textcolor{black}{The \ac{mpnn} takes as input the pole measurements $\mathbf{z}_{v,t}^{\text{(O)}}$ and $\boldsymbol{\rho}_{v,t}^{\text{(O)}}$, and constructs the input graph $\mathcal{G}$ as detailed in Sec.~\ref{subsec:data_association_model}.
The aim of the \ac{mpnn} is to learn the correct association among the pole measurements by operating on the graph $\mathcal{G}$ with an iterative message passing scheme.}
The overall process is composed by $I$ message passing iterations and for each iteration $i~=~1,\dots, I$ the node and edge embeddings, denoted as $\mathbf{h}_n^{(i)}$ and  $\mathbf{m}_{n\to m}^{(i)}$, are used to spread the information throughout $\mathcal{G}$.
More specifically, at iteration $i = 0$, the node and edge embeddings are initialized as
\begin{align}
\mathbf{m}_{n\to m}^{(0)} &= g_{\text{e}}^{\text{enc}}\left(\overline{\mathbf{z}}_{v,k,1,t}^{\text{(O)}} - \overline{\mathbf{z}}_{v',k',1,t}^{\text{(O)}} , \overline{\mathbf{z}}_{v,k,8,t}^{\text{(O)}} - \overline{\mathbf{z}}_{v',k',8,t}^{\text{(O)}}\right) \, , \nonumber \\ 
& \hspace{-25pt} \forall (n,m) \in \mathcal{E}: \Phi_t(n) = \{k,v\} \land \Phi_t(m) = \{k',v'\}, v \neq v' ,
\\
\mathbf{h}_{n}^{(0)} &= g_{\text{n}}^{\text{enc}}\left(\overline{\mathbf{z}}_{v,k,t}^{\text{(O)}}\right) \,, \quad \forall n \in \mathcal{V}: \Phi_t(n) = \{k,v\},
\end{align}
where, $g_{\text{n}}^{\text{enc}}(\cdot)$ and $g_{\text{e}}^{\text{enc}}(\cdot)$ are two \acp{mlp} at the node and edge, respectively, whose goal is to extract the feature embeddings from the bounding boxes.
The \ac{mpnn} then further encodes the node and edge embeddings using three additional \acp{mlp} at each node, namely $g_{\text{n}}(\cdot)$, $g_{\text{n}}^{\text{out}}(\cdot)$ and $g_{\text{n}}^{\text{in}}(\cdot)$, while an additional \ac{mlp} $g_{\text{e}}(\cdot)$ is used at each edge.
The role of each \ac{mlp} is detailed in~\cite{TedBraBarNic:C22}. 
At iteration $i$, the \ac{mlp} at each edge is  used to update the edge embeddings as 
\begin{equation}
	\label{edge_update_classic}
	\mathbf{m}_{n\to m}^{(i)} = g_{\text{e}}\left(\mathbf{h}_n^{(i-1)}, \mathbf{h}_m^{(i-1)}, \mathbf{m}_{n\to m}^{(i-1)}\right) \,,  \ \forall m~\in~\mathcal{N}_n \,,
\end{equation}
where $\mathcal{N}_n $ is the set of neighbors of node $n~\in~\mathcal{V}$, while the node embeddings are updated as 
\begin{equation}
	\label{node_update_mpnn}
	\mathbf{h}_n^{(i)} = g_{\text{n}}\left(\sum_{ m~\in~\mathcal{N}_n}  \mathbf{m}_{m\to n}^{\text{in},(i)}, \sum_{ m~\in~\mathcal{N}_n}  \mathbf{m}_{n\to m}^{\text{out},(i)}\right),
\end{equation}
where 
\begin{align}
	\label{node_update_mpnn_in}
	\mathbf{m}_{m\to n}^{\text{in},(i)} &= g_{\text{n}}^{\text{in}}\left(\mathbf{h}_i^{(i-1)}, \mathbf{m}_{m\to n}^{(i)}\right)\,, \quad & \forall m~\in~\mathcal{N}_n,
\end{align}
and
\begin{align}
	\label{node_update_mpnn_out}
	\mathbf{m}_{n\to m}^{\text{out},(i)} &= g_{\text{n}}^{\text{out}}\left(\mathbf{h}_n^{(i-1)}, \mathbf{m}_{n\to m}^{(i)}\right)\,, \quad & \forall m~\in~\mathcal{N}_n.
\end{align}
are the incoming and outgoing edge embeddings, respectively.

This procedure is repeated for $I$ message passing iterations and the association variable $\widehat{\mathtt{a}}_{n \to m}$ is obtained by applying a thresholding operation (with threshold $\Gamma$) to the output of an edge classifier, called $g_{\text{e}}^{\text{class}}(\cdot)$, operating over the edge embeddings as 
$\widehat{\mathtt{a}}_{n\to m}^{(I)} = g_{\text{e}}^{\text{class}}\left(\mathbf{m}_{n\to m}^{(I)}\right) \geq \Gamma $.
\textcolor{black}{Other strategies can be used for obtaining $\widehat{\mathtt{a}}_{n\to m}^{(I)}$, such as the Hungarian algorithm.}
\textcolor{black}{Table \ref{tab:mpnn_arch} summarizes the layers used by each encoder in the \ac{mpnn} architecture. 
Note that after each linear layer, a dropout operation and a ReLU activation function are added.}

\textcolor{black}{The \ac{mpnn} is trained using a weighted binary cross-entropy loss computed after each  
iteration $i$ as
\begin{align}
     \label{eq:loss}
   \mathcal{L}_{\text{MPNN}} =& \frac{-1}{|\mathcal{E}|} 
   \sum_{i = 1}^{I}  
      \sum_{(n,m)~\in~\mathcal{E}}  \!\!\!\!
     (1-{\mathtt{a}}_{n\to m})
     \text{log}(1 - \widehat{\mathtt{a}}_{n\to m}^{(i)})     \nonumber \\  & + w \, {\mathtt{a}}_{n\to m}\text{log}(\widehat{\mathtt{a}}_{n\to m}^{(i)})\,,
\end{align}
where $w$ is a weight given to the positive class to compensate for the class unbalances, which is computed as the ratio between the number of 0-class and 1-class edges.}

\begin{table*}[!t]
	\renewcommand{\arraystretch}{1.3}
	\setlength{\tabcolsep}{2.5pt}

 \caption{\textcolor{black}{MPNN architecture}}
	\label{tab:mpnn_arch}
	\centering
 \vspace{-2mm}
	\begin{tabular}{|c|c||c|c||c|c||c|c||c|c||c|c|} \hline
 \multicolumn{2}{|c||}{\textcolor{black}{$g_n^{(enc)}$}} & \multicolumn{2}{c||}{\textcolor{black}{$g_e^{(enc)}$}} & 
 \multicolumn{2}{c||}{\textcolor{black}{$g_n$}} &
 \multicolumn{2}{c||}{\textcolor{black}{$g_e$}} & 
 \multicolumn{2}{c||}{\textcolor{black}{$g_e^{(class)}$}} & \multicolumn{2}{c|}{\textcolor{black}{$g_e^{(in)} = g_e^{(out)}$}} \\ \hline
    \textcolor{black}{\textbf{Layer type}} & \textcolor{black}{\textbf{Output}} & \textcolor{black}{\textbf{Layer type}} & \textcolor{black}{\textbf{Output}} & 
    \textcolor{black}{\textbf{Layer type}} & \textcolor{black}{\textbf{Output}} & 	
    \textcolor{black}{\textbf{Layer type}} & \textcolor{black}{\textbf{Output}} & 	
    \textcolor{black}{\textbf{Layer type}} & \textcolor{black}{\textbf{Output}} & 	
    \textcolor{black}{\textbf{Layer type}} & \textcolor{black}{\textbf{Output}} \\ \hline\hline 
    \textcolor{black}{Input} & \textcolor{black}{$24 \times 1$} & \textcolor{black}{Input} & \textbf{\textcolor{black}{$6 \times 1$}} & \textcolor{black}{Input} & \textbf{\textcolor{black}{$64 \times 1$}} & 
    \textcolor{black}{Input} & \textbf{\textcolor{black}{$96 \times 1$}} & \textcolor{black}{Input} & \textbf{\textcolor{black}{$16 \times 1$}} & \textcolor{black}{Input} & \textbf{\textcolor{black}{$48 \times 1$}} \\ 
    \textcolor{black}{Linear} & \textbf{\textcolor{black}{$72 \times 1$}} & \textcolor{black}{Linear} & \textbf{\textcolor{black}{$18 \times 1$}} & \textcolor{black}{Linear} & \textbf{\textcolor{black}{$32 \times 1$}} & \textcolor{black}{Linear} & \textbf{\textcolor{black}{$80 \times 1$}} & 
    \textcolor{black}{Linear} & \textbf{\textcolor{black}{$8 \times 1$}} & \textcolor{black}{Linear} & \textbf{\textcolor{black}{$56 \times 1$}} \\
     \textcolor{black}{Linear} & \textbf{\textcolor{black}{$32 \times 1$}} & \textcolor{black}{Linear} & \textbf{\textcolor{black}{$18 \times 1$}} & & &  \textcolor{black}{Linear} & \textbf{\textcolor{black}{$16 \times 1$}} & \textcolor{black}{Sigmoid} & \textbf{\textcolor{black}{$1 \times 1$}} & \textcolor{black}{Linear} & \textbf{\textcolor{black}{$32 \times 1$}} \\ 
     & & \textcolor{black}{Linear} & \textbf{\textcolor{black}{$16 \times 1$}} & & & & & & & & \\ \hline 
	\end{tabular}
 \vspace{-11pt}
\end{table*}



\subsection{Measurement-to-target association}
\label{subsec:meas_to_target_association}

Once the association variable $ \widehat{\mathtt{a}}_{n\to m}$ is estimated, we can distinguish which set of bounding boxes from different vehicles have been generated by the same object.
This is done by extracting the connected components of the graph $\mathcal{G}$ with only the positive-classified edges, i.e., $\widehat{\mathtt{a}}_{n\to m}^{(I)}>\Gamma$. We define with $\mathcal{A}_{\widehat{o},t} = \{(v,k): \Phi_t(n) = \{k,v\} \land \widehat{\mathtt{a}}_{n\to m}^{(I)}>\Gamma\}$ the set of vehicle-measurements tuples that are related to the same target with unknown identification $\widehat{o}$.  The goal now is to associate each $\widehat{o}$ to the indexes  $o \in \mathcal{O}_{t-1}$.

From the mean of the associated bounding boxes, we estimate the object position by computing the centroid averaging over the eight corners as
\begin{equation}
	\widehat{\mathbf{x}}_{\widehat{o},t}^{\text{(O)}}
	= 
	\dfrac{1}{8 \, |\mathcal{A}_{\widehat{o},t}| }
	\sum_{c=1}^{8} 
	\sum_{(v,k)\in \mathcal{A}_{\widehat{o},t}} \overline{\mathbf{z}}_{v,k,c,t}^{(O)}  \,,  \quad \forall \,\widehat{o}. 
\end{equation}

Motivated by the fact that $ \widehat{\mathbf{x}}_{\widehat{o},t}^{\text{(O)}}$ holds a good approximation of the real bounding boxes positions, we solve this second association step by setting  $\alpha_{v,o,t} = 1$ according to the maximum likelihood criterion over the  distance  $d_{\widehat{o},o,t} =\| \widehat{\mathbf{x}}_{\widehat{o},t}^{\text{(O)}} -   \widehat{\mathbf{x}}_{o,t-1}^{(\text{O})} \|_2$, constrained to $d_{\widehat{o},o,t}< \Delta $, being $\Delta$ a given threshold. Otherwise, $\alpha_{v,o,t} = 0$ indicates the detection of a newly-observed object.

\section{\textcolor{black}{Implementation aspects}}
\label{sec:challenges}

\textcolor{black}{This section highlights potential challenges that can arise in the implementation of the proposed cooperative approach in real systems. Specifically, we discuss about the limitations of LiDAR sensors in
Sec.~\ref{subsec:challenges_lidar_sensing}, while in Sec.~\ref{subsec:challenges_synchr} we detail strategies to limit the impact of inaccurate synchronization, network reliability and signal interference. Privacy and security issues
are covered in Sec.~\ref{subsec:privacy_considerations}. }

\subsection{\textcolor{black}{LiDAR sensing technology limitations}}
\label{subsec:challenges_lidar_sensing}

\textcolor{black}{Despite the highly-accurate environmental sensing, LiDAR sensors are impaired by noise, occlusion, and adverse weather conditions that require specific countermeasures.
For example, \ac{ml}-based methodologies could be designed to reduce the point cloud noise (see e.g.,~\cite{lidar_denoising1}).
Similarly, data-driven strategies could be utilized to predict objects that are partially occluded (see e.g.,~\cite{occlusion}), reducing the degradation due to blockage. 
Under rain, fog, or snow conditions, the environmental sensing obtained by LiDAR may not be sufficiently accurate, and the integration of multiple imaging sensors (e.g., camera, radar, and LiDAR) is suggested to provide a reliable localization system under all weather conditions (for example, by drawing inspiration from~\cite{fusion_imaging}).}

\subsection{\textcolor{black}{ Synchronization, network reliability and signal interference}}
\label{subsec:challenges_synchr}


\textcolor{black}{The performance of our cooperative localization system can be degraded by the impairments of the communication channel such as poor synchronization, communication delays, unreliable networks, and signal interference. 
All these aspects need to be handled by dedicated protocols or processing strategies.
In particular, synchronization between vehicles and the road infrastructure could be guaranteed by dedicated protocols, such as the \ac{ptpv2}~\cite{scientific_reports_bernardo}, or by using synchronization signals extracted from the onboard \ac{gnss} receivers of the vehicles.
Communication delays introduced by the network must be handled by temporally realigning the measurements received by the infrastructure as discussed in~\cite{icp}. Alternatively, the asynchronicity of measurements can be statistically accounted for within the tracking algorithm~\cite{Gaglione22}.}
\textcolor{black}{For what concerns unreliable networks, vehicles could implement retransmission schemes or the infrastructure could modify the fusion strategy to take into account packet drops.}   
\textcolor{black}{Signal interference among connected vehicles could be reduced by introducing ad-hoc scheduling strategies~\cite{scheduling_v2i} or resource allocation policies~\cite{resource_allocation_v2i}. 
}

\subsection{\textcolor{black}{Privacy and security considerations}}
\label{subsec:privacy_considerations}

\textcolor{black}{From a privacy and security perspective, the proposed approach could leak information about the area where  vehicles are currently located.
This may be inferred from the information sent by the vehicles to the \ac{mec}. 
Additionally, unauthorized entities are able to tamper with the communications to modify the exchanged vehicles and/or object positions, jeopardizing the \ac{cls-mpnn} positioning performances. 
All these situations must be avoided by introducing ad-hoc security schemes.
For example, the communication can be encrypted by adopting blockchain technologies as discussed in~\cite{blockchain}.  
On the other hand, the impact of malicious vehicles entering the cooperative process can be limited by employing authentication protocols~\cite{auth_protocols}. 
}

\section{Neural networks training and evaluation}
\label{sec:nn_results}

\textcolor{black}{This section details the strategies adopted for training the LiDAR-based 3D object detectors and the \ac{mpnn} and their respective performances. 
In particular, Sec.~\ref{subsec:obj_detection_results} provides the results obtained by the 3D object detectors, while Sec~\ref{subsec:mpnn_results} focuses on the \ac{mpnn} \ac{da} performances.}



\subsection{LiDAR-based object detection}
\label{subsec:obj_detection_results}


This section focuses on the evaluation of the detection performances of PointPillars and Part-$\text{A}^2$.
We are interested in characterizing their ability to correctly estimate the bounding boxes of the poles from the LiDAR point clouds to evaluate the overall error introduced by the detection process while assuming the vehicle position to be exactly known.

For training the 3D object detectors, we generated a synthetic dataset from CARLA~\cite{pmlr-v78-dosovitskiy17a}, an advanced  driving simulator that allows flexible configuration of realistic vehicular layouts and sensor specifications.
The dataset is collected in the urban map \textit{Town02} of CARLA, that spans roughly a $200\!\times\! 200$~$\text{m}^{2}$ area, where a single LiDAR-equipped vehicle moves in the environment for 5000 timestamps sampled at $T_s=0.2$~s.

The LiDAR sensor is placed on the roof of the vehicle and is configured to have 64 planes covering an elevation angle between $-25$° and $15$°, a sensing range of $70$ m and to output 1 million point clouds per second, on average, with a $\pm 2$~cm accuracy. 
For a more realistic simulation, we randomly remove $20\%$ of the points for each captured frame and apply an additional dropping strategy where points having a reflectance value $r_j < 0.8$ are randomly dropped. 
The ground truth data consists of 3D bounding boxes physically enclosing the poles for each captured LiDAR frame.
Note that we consider only ground-truth boxes with at least 10 points inside.  
The performances of the object detector are evaluated over a held-out portion of the dataset composed by $1000$ examples, while the remaining $4000$ ones are used for training.

\textcolor{black}{Both detectors are implemented relying on the open-source repository in~\cite{openpcdet}.
PointPillars is configured to have $\mathcal{P}= 16000$ pillars of $N_p = 32$ points each, while the output image dimensions obtained after scattering back the learned features are set to $C=64$, $H=432$ and $W=496$.
On the other hand, Part-$\text{A}^2$ utilizes 16000 voxels.
The prior boxes width, length, and height are selected as  1.1~m, 1.1~m, and 7.67~m, respectively, for both detectors.
PointPillars is trained over $160$ epochs with batches of $6$ examples using an Adam optimizer~\cite{adam} with learning rate $\eta = 0.01$, momentum parameters $\beta_1 = 0.95$ and $\beta_2 = 0.85$, and a weight decay factor of 0.01.
A cosine annealing strategy~\cite{cosine_annealing} without warm restarts is used to schedule the learning rate.
Similarly, Part-$\text{A}^2$ is trained using the same optimizer and learning rate scheduler but considering a maximum number of epochs of $80$ and with a batch size of $2$ examples.}

\textcolor{black}{Data augmentation strategies are used to improve the performances of both detectors. 
During training, each point cloud is augmented by applying a random rotation of [-90°, +90°], a random scaling of [0.95,1.05], and a random horizontal flip.
Additionally, ground-truth sampling \cite{second} is used, where at most $10$ bounding boxes are added to each point cloud example. 
For Part-$\text{A}^2$, an additional augmentation strategy is adopted, whereby the ground-truth sampling is applied by adding noise to the bounding boxes. Specifically, two types of noise are considered: a translational noise which moves the bounding box on the horizontal plane, and a rotational noise, which adds a random rotation in the interval [-45°, +45°].}



As performance metric we use the \ac{ap} as presented in~\cite{pascal_voc}.
For object detection tasks, the \ac{ap} requires the computation of the \ac{iou} to discern positive and negative predictions.
Given a pair $\{\mathbf{b}_j, \widehat{\mathbf{b}}_j \}$ of ground-truth and predicted bounding boxes, the \ac{iou} is the ratio of the volume of the intersection and the volume of the union between $\mathbf{b}_{j}$ and $\widehat{\mathbf{b}}_{j}$.
Thus, the \ac{iou} is
\begin{equation}
    \text{IoU}(\mathbf{b}_j, \widehat{\mathbf{b}}_j)~=~\dfrac{\text{volume}(\mathbf{b}_j \cap \widehat{\mathbf{b}}_j)}{\text{volume}(\mathbf{b}_j \cup \widehat{\mathbf{b}}_j)} \,,
\end{equation}
and $\widehat{\mathbf{b}}_{j} $ is deemed as positive if the \ac{iou} is greater than a threshold $\varepsilon$.
The \ac{ap} is evaluated considering two recall levels $L=11$ and $L~=~40$ as recently suggested in~\cite{map_recall_levels}.

As an additional performance metric, we evaluate the localization error of the predicted bounding boxes with respect to the ground truths, neglecting any \acp{fp}.
We consider the centroid of the bounding box to assess how accurate and precise are the predictions with respect to the ground truths.
For this purpose, we evaluate the \ac{cep} at $9$5\% confidence of the location error \ac{cdf} over each axis separately.
Both performance metrics are evaluated on the validation dataset. 


\begin{table*}[!t]
	\renewcommand{\arraystretch}{1.3}
	\setlength{\tabcolsep}{2.5pt}
	\caption{3D object detection performance}
	\label{tab:map}
	\centering
	\begin{tabular}{|c|c|c|c|c|c|c|c|c|c|c|c|c|}
		\hline
		\multirow{3}{*}{\textbf{Detector}}  & \multicolumn{6}{c|}{\textbf{$\text{AP}_{\text{BEV}}$}} &  \multicolumn{6}{c|}{\textbf{$\text{AP}_{\text{3D}}$}} \\ \cline{2-13}
		& \multicolumn{3}{c|}{$L~\!=\!~11$} & \multicolumn{3}{c|}{$L~\!=\!~40$} & \multicolumn{3}{c|}{$L~\!=\!~11$} & 
		\multicolumn{3}{c|}{$L~\!=\!~40$} \\  \cline{2-13}
		& $\varepsilon~\!=\!~0.25$ & $\varepsilon~\!=\!~0.50$ & $\varepsilon~\!=\!~0.70$ & $\varepsilon~\!=\!~0.25$ & $\varepsilon~\!=\!~0.50$ & $\varepsilon~\!=\!~0.70$ & $\varepsilon~\!=\!~0.25$ & $\varepsilon~\!=\!~0.50$ & $\varepsilon~\!=\!~0.70$ & $\varepsilon~\!=\!~0.25$ & $\varepsilon~\!=\!~0.50$ & $\varepsilon~\!=\!~0.70$ \\ \hline
		PointPillars & 97.27 & 97.21 & 90.35 & 98.75 & 98.61 & 95.99 & 97.16 & 96.24  & 89.35 & 98.74 & 98.32 & 95.45 \\
		Part-$\text{A}^2$ & 89.54 & 89.41 & 75.64 & 94.49 &  92.31 & 76.10 & 89.49 & 89.22 & 75.08 & 92.67 & 92.38 & 75.96\\ \hline
	\end{tabular}
 \vspace{-11pt}
\end{table*}


Table~\ref{tab:map} reports the \ac{ap} for PointPillars and Part-$\text{A}^2$ considering \ac{iou} thresholds $\varepsilon~=~\{0.25, 0.50,0.70\}$ and recall levels $L~=~\{11, 40\}$.
The \ac{ap} is evaluated over the \ac{bev} projection, denoted as $\text{AP}_{\text{BEV}}$, which projects the predictions onto the $x$-$y$ plane and in the 3D space, referred to as $\text{AP}_{\text{3D}}$.
Comparing the results, the $\text{AP}_{\text{BEV}}$ and $\text{AP}_{\text{3D}}$ metrics show that PointPillars provides more accurate detections compared to Part-$\text{A}^2$ for all values of $L$ and \ac{iou} thresholds $\varepsilon$. 
Indeed, the \ac{ap} of PointPillars is, on average, $5\%$ higher than the one obtained by  Part-$\text{A}^2$ until $\varepsilon \leq 0.50$, while for more stringent thresholds, i.e., $\varepsilon~=~0.7$, the gap between the two detectors becomes even larger.
Focusing on the results under different recall levels $L$, one can notice that $L~=~11$ provides a lower value compared to $L~=~40$.
This should be expected since $L~=~11$ generally leads to an underestimation of the actual performances compared to $L~=~40$ as relying only on 11 recall values compared to 40.

\begin{figure}[!t]
\centering
	\centering
	\includegraphics[width=0.95\columnwidth]{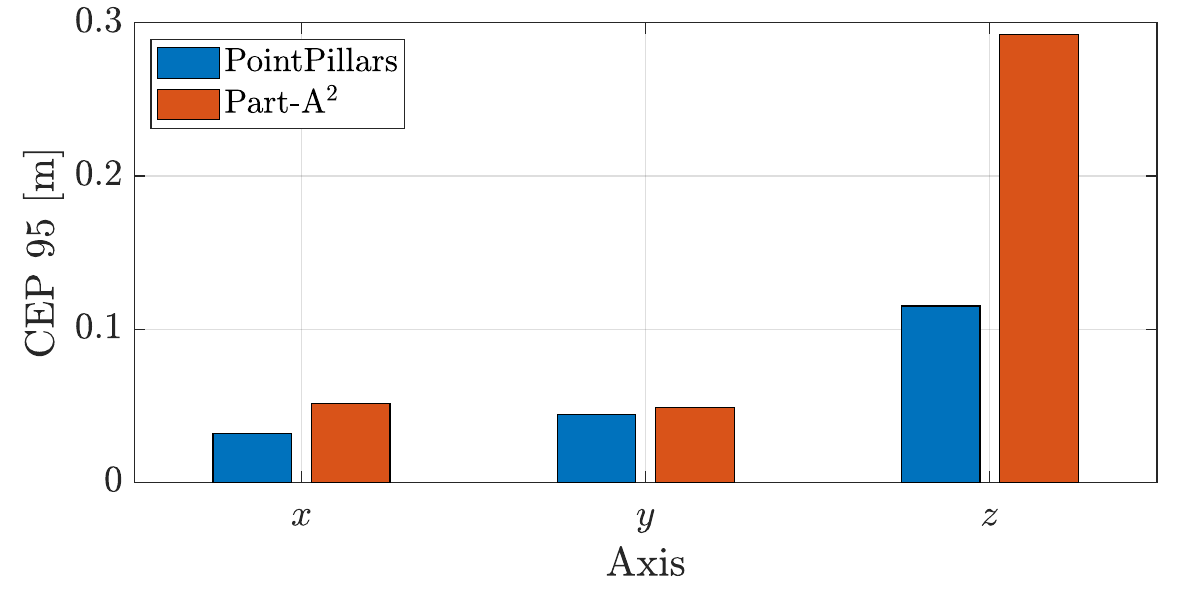}
    \vspace{-4mm}
    \caption{Bounding box location error CEP95 over the three axes.}
    \label{fig:bbox_location_cdf}
    \vspace{-5mm}
\end{figure}

To further characterize the performance of the 3D object detectors, in Fig.~\ref{fig:bbox_location_cdf} we report the \ac{cep}95 of the location error obtained by PointPillars and Part-$\text{A}^2$ over the three axes.
The results confirm the findings of the previous analysis: PointPillars provides more accurate detections when compared with Part-$\text{A}^2$.
More specifically, the two methods have similar \ac{cep}95 over the $x$ and $y$ axes, while for the $z$ axis PointPillars achieves a location error lower than $13$~cm in 95\% of the cases, while Part-$\text{A}^2$ attains a much larger value of $29$ cm.

\subsection{MPNN data association}
\label{subsec:mpnn_results}

The \ac{mpnn} is trained over the same CARLA map used for training the 3D object detectors.
In particular, a new dataset is generated in \textit{Town02} where 20 LiDAR-equipped vehicles, with LiDAR sensors configured as in Sec.~\ref{subsec:obj_detection_results}, move in the environment for $1000$ timestamps, again with sampling time $T_s~=~0.2$s. 
The database is constructed by firstly running the PointPillars model on the point cloud data acquired by each vehicle individually to obtain the set of estimated bounding boxes.
We specifically select PointPillars in this stage due to its more accurate detection performances compared to Part-$\text{A}^2$ as highlighted in the previous analysis.
\textcolor{black}{To make the \ac{mpnn}-based \ac{da} strategy more resilient to the errors discussed in Sec.~\ref{subsec:data_association_model}, we add Gaussian noise to the bounding boxes provided by PointPillars.
The noise standard deviation is set to $1$~m.}
Then, the input graphs containing the true associations among the poles detected by the vehicles are used as ground-truth information for training the \ac{mpnn}.
We use $700$ examples for training and the remaining $300$ for validation. 
\textcolor{black}{The MPNN is trained considering a batch size of $20$ examples for 50 epochs and using the Adam optimizer with learning rate $\eta = 0.001$ and momentum parameters $\beta_1 = 0.9$ and $\beta_2 = 0.999$.}

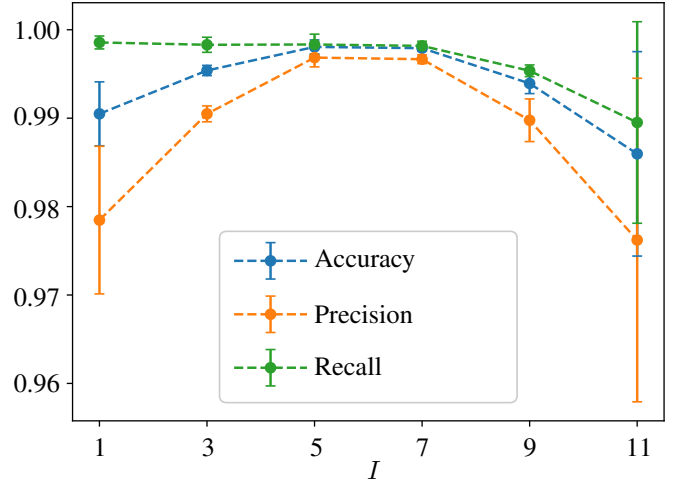
\begin{figure}[!tb]
    \centering
    \begin{tikzpicture}
        \node[]at(0,0){\includegraphics[width=0.9\columnwidth , keepaspectratio]{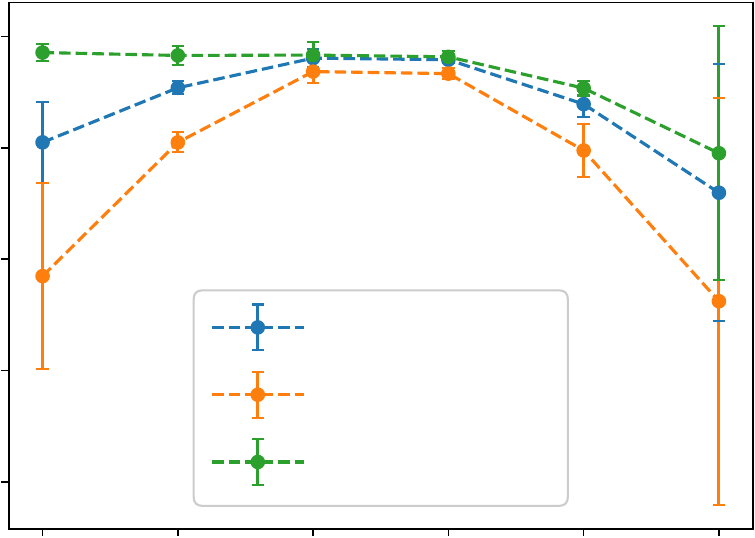}};

        \node[right]at(-4.8,2.45){1.00};
        \node[right]at(-4.8,1.3){0.99};
        \node[right]at(-4.8,0.07){0.98};
        \node[right]at(-4.8,-1.1){0.97};
        \node[right]at(-4.8,-2.25){0.96};
        
        \node[right]at(3.34,-3.1){11};
        \node[right]at(1.98,-3.1){9};
        \node[right]at(0.55,-3.1){7};
        \node[right]at(-0.86,-3.1){5};
        \node[right]at(-2.3,-3.1){3};
        \node[right]at(-3.74,-3.1){1};
        
        \node[right]at(-0.1,-3.4){$I$};
        
        \node[right]at(-0.8,-0.625){Accuracy};
        \node[right]at(-0.8,-1.325){Precision};
        \node[right]at(-0.8,-2.025){Recall};
    \end{tikzpicture}
   \vspace{-8mm}
   \protect\caption{\ac{da} task validation performances reached after 50 epochs of training as a function of the number of message passing iterations $I$.}
    \label{fig:T_choice}
    \vspace{-4mm}
\end{figure}

To tune the hyper-parameter $I$, we analyze the accuracy, precision, and recall performance metrics in the validation dataset by varying the number of message passing iterations, whose results are reported in Fig.~\ref{fig:T_choice}. The error bars are computed using the mean values and standard deviation as uncertainties. Results show that the optimal number of iterations $I$ is between 5 and 7. Indeed, $I < 5$ prevents effective feature extractions and elaboration, while $I > 7$ may lead to over-complicated models and potential overfitting. That is, when $I$ is too large, the model might start learning noise or outliers in the training data, rather than the underlying patterns.
\textcolor{black}{Therefore, for the following analyses we set $I = 5$.}

\section{Cooperative localization results}
\label{sec:coop_loc_results}

This section analyzes the performances of the proposed cooperative localization approach in two realistic driving scenarios.
\textcolor{black}{The goal is to comprehensively assess the DA approach and the overall cooperative system as compared to ideal and state-of-the-art solutions.}
The first one is again based on the map \textit{Town02} of CARLA but with new trajectories and vehicle interactions (Sec. \ref{subsec:town02_results}). 
The second one instead is a brand new environment where the \acp{nn} have been not trained on (Sec. \ref{subsec:town10_results}).
\textcolor{black}{Finally, Sec.\ref{subsec:communication_overhead} discusses the communication overhead of \ac{cls-mpnn}.}


\subsection{Town02 results}
\label{subsec:town02_results}

We initially assess the proposed method considering the \textit{Town02} map of CARLA where twenty vehicles move in the area following different trajectories compared to the ones of Sec. \ref{sec:nn_results}.
\textcolor{black}{Unless stated otherwise, we assume that the onboard \ac{gnss} receiver provides a position estimate with accuracy $\sigma_{\text{P}}^{(\text{V})} = 5$~m all over the map.}
The same parameters used in Sec.~\ref{sec:nn_results} are reused to configure the LiDAR sensors onboard the vehicles. 
The motion model \eqref{eq:motion_model_vehicle} is calibrated according to the vehicles' motion, with $\sigma_{\text{Q}} = 0.5$  m/${\text{s}^2}$, while the accuracy of the object detectors is as in Fig.~\ref{fig:bbox_location_cdf}.
The overall simulation lasts for $300$ seconds and comprises $1500$ snapshots each sampled at $T_s = 0.2$ s. 
\begin{table}[!t]
\vspace{-3mm}
	\renewcommand{\arraystretch}{1.3}
	\setlength{\tabcolsep}{4pt}
	\caption{Detection Performances Analysis in Town02}
	\label{tab:detection_town02}
	\centering
 \vspace{-2mm}
	\begin{tabular}{|c|c|c|c|}
		\hline
		\multirow{2}{*}{\textbf{Method}}  & \textbf{Correct detections} & \textbf{Missed detections} & \textbf{False positives} \\
		& [\%] & [\%] & [\%] \\ \hline
		PointPillars & 95.93 & 4.07 & 12.59 \\
		Part-$\text{A}^2$ & 94.30 & 5.70 & 16.72 \\ \hline
	\end{tabular}
  \vspace{-4mm}
\end{table}

\begin{figure*}
	\centering
	\subfloat[\label{fig:rmse_map_gps_town02}]{
		\includegraphics[width=0.32\linewidth]{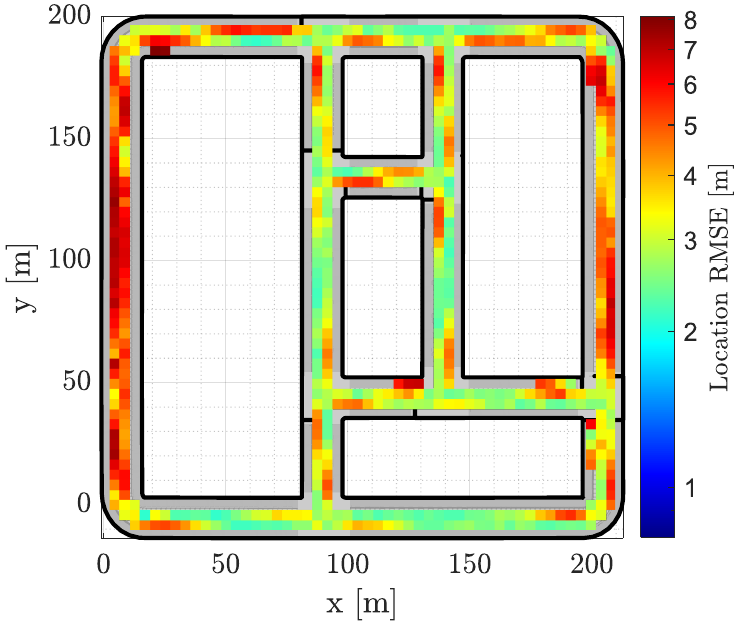}
	}
	\subfloat[\label{fig:rmse_map_icp_oracle_town02}]{
		\includegraphics[width=0.32\linewidth]{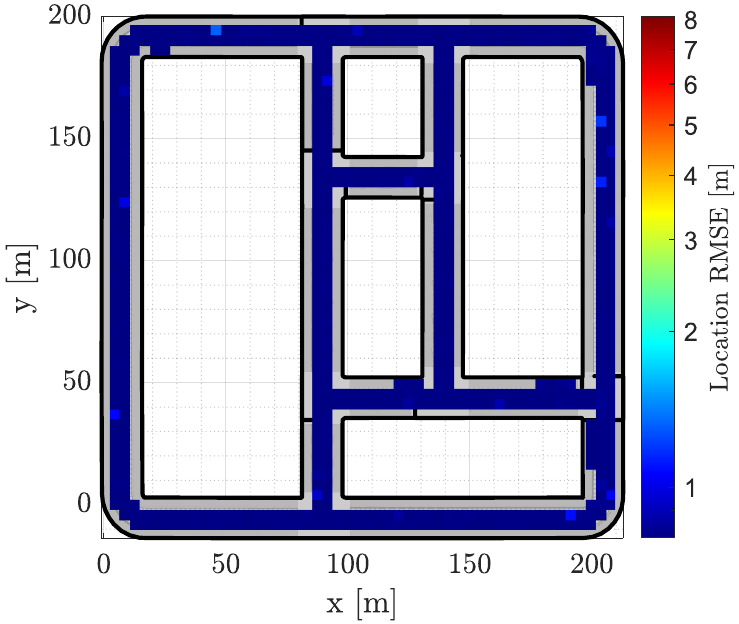}
	}
	\subfloat[\label{fig:rmse_map_icp_pointpillars_town02}]{
		\includegraphics[width=0.32\linewidth]{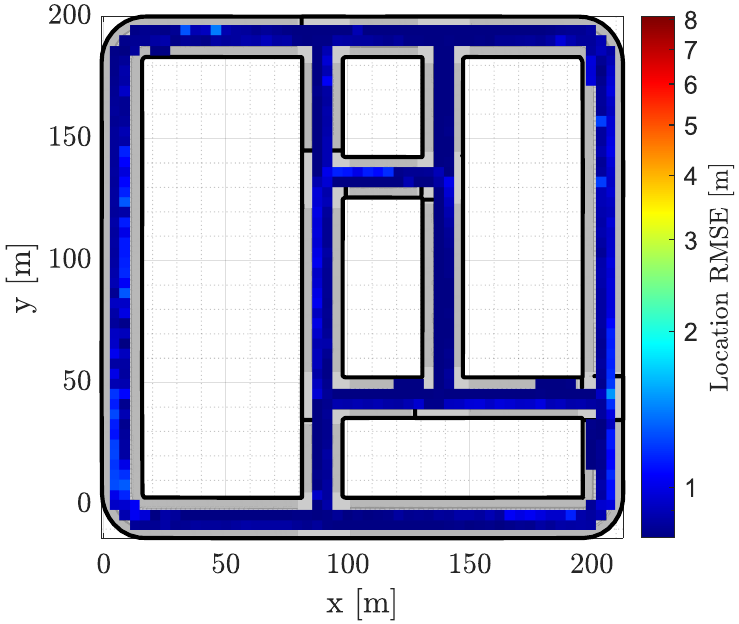}
	}
	
	\vspace{-4mm}
	\subfloat[\label{fig:rmse_map_icp_parta2_town02}]{
		\includegraphics[width=0.32\linewidth]{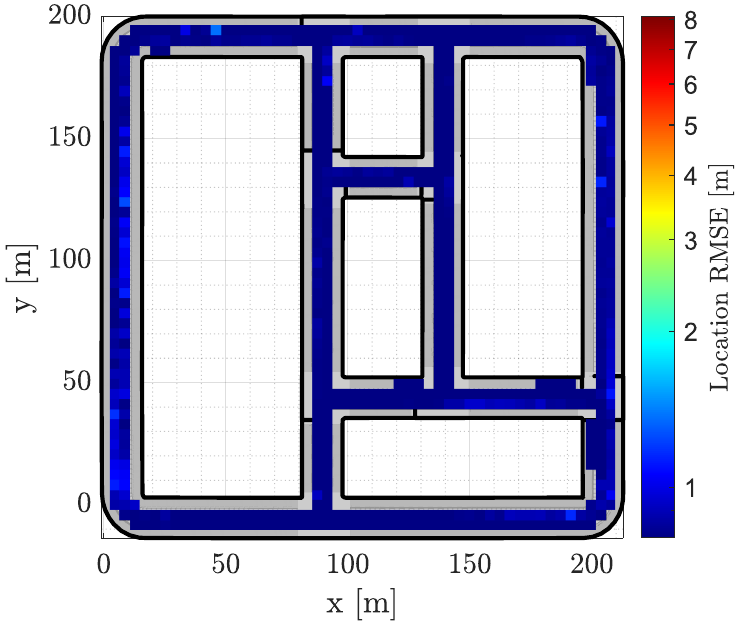}
	}
	\subfloat[\label{fig:rmse_map_slam_town02}]{
		\includegraphics[width=0.32\linewidth]{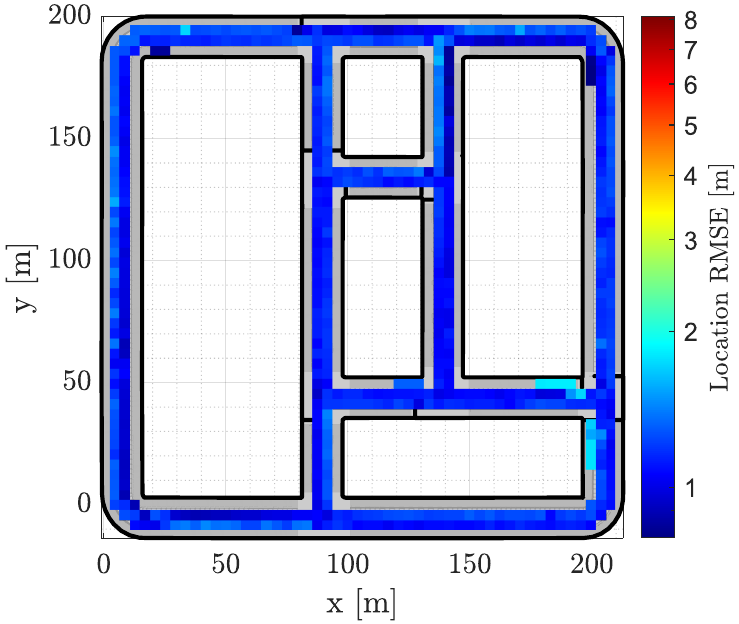}
	}	
	\subfloat[\label{fig:rmse_comparison_town02}]{
		\hfill
		\includegraphics[width=0.32\linewidth]{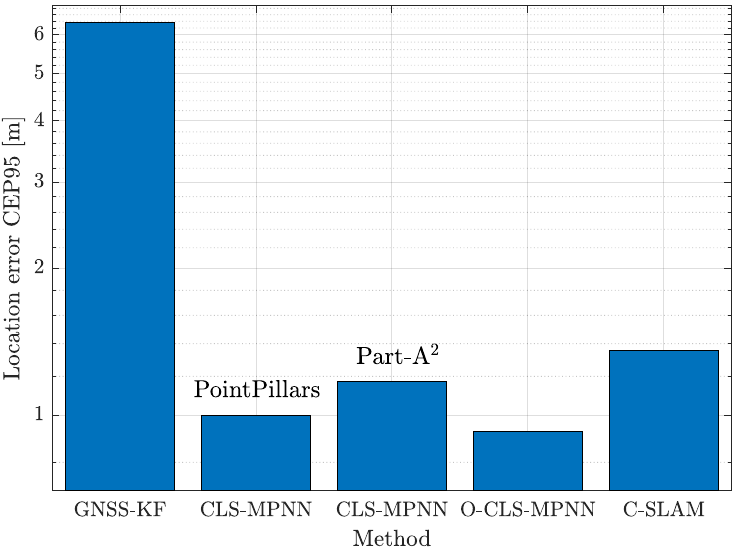}
	}
	\vspace{-1mm}
	\caption{\textcolor{black}{Spatial RMSE of vehicle positions over the \textit{Town02} simulation map:  (a) GNSS-KF. (b) \ac{o-cls-mpnn} (perfect \ac{da} and LiDAR sensing) (c) \ac{cls-mpnn} with PointPillars, (d) \ac{cls-mpnn} with Part-$\text{A}^2$, (e) C-SLAM, (f) comparison of the CEP95 of the vehicle location error for all the methodologies.}}
	\label{fig:results_comparison_town02}
	\vspace{-4mm}
\end{figure*}

In the following, we assess the performances provided by the following positioning solutions:
\begin{itemize}
    \item The proposed \ac{cls-mpnn} tool integrating the detections provided by the 3D object detectors and the \ac{mpnn}-based \ac{da} strategy. We will show results considering separately the detections obtained by PointPillars and Part-$\text{A}^2$.  
	\item An oracle version of the proposed method, referred to as \ac{o-cls-mpnn}, with perfect \ac{da} and LiDAR sensing where all poles present in the field of view of the LiDAR sensor are detected.
	\item A state-of-the-art cooperative \ac{slam} approach~\cite{9775023} that requires vehicles to exchange their positions and the raw LiDAR point clouds, denoted as C-SLAM.
	\item A non-cooperative \ac{gnss}-based tracking tool where a \ac{kf} integrates the positions provided by the \ac{gnss} units, referred to as GNSS-KF.  
\end{itemize} 
The main difference between \ac{o-cls-mpnn} and \ac{cls-mpnn} is that \ac{cls-mpnn}  is subject to missed detections while \ac{o-cls-mpnn} is not as all detectable poles are always correctly recognized, provided they fall inside the field of view of the LiDAR sensor. 
The detection performances of PointPillars and Part-$\text{A}^2$ are reported in Table \ref{tab:detection_town02} which highlights the percentages of correct detections, missed detections, and false positives obtained by the two detectors. 
The results indicate that PointPillars provides more robust results as its missed detection and false positive percentages are lower when compared with the ones attained by Part-$\text{A}^2$.

\subsubsection{\textcolor{black}{Cooperative localization performances}}
\label{subsubsec:coop_loc_results_town02}
\textcolor{black}{Fig.~\ref{fig:results_comparison_town02} shows the spatial \ac{rmse} of the vehicle location estimate computed all over the map for GNSS-KF (Fig.~\ref{fig:rmse_map_gps_town02}), \ac{o-cls-mpnn} (Fig.~\ref{fig:rmse_map_icp_oracle_town02}), the proposed \ac{cls-mpnn} method considering the detections provided by PointPillars (Fig.~\ref{fig:rmse_map_icp_pointpillars_town02}), Part-$\text{A}^2$ (Fig.~\ref{fig:rmse_map_icp_parta2_town02}) and C-SLAM (Fig.~\ref{fig:rmse_map_slam_town02}).
Additionally, Fig.~\ref{fig:rmse_comparison_town02} reports the \ac{cep}95 of the vehicle location error for all the compared methodologies.}
Analyzing the results, \ac{cls-mpnn} substantially outperforms GNSS-KF on all map positions, regardless of the detector employed, while approaching the performances attained by \ac{o-cls-mpnn}.
\textcolor{black}{The improvement brought by \ac{cls-mpnn} is nearly constant all over the map apart from a few spots in the leftmost road where a decreased level of cooperation is observed due to a lower number of vehicles driving in that area.}
The developed method also provides more accurate results compared to the ones achieved by C-SLAM as its location \ac{rmse} is largely reduced both for PointPillars and Part-$\text{A}^2$. 
Looking at the \ac{cep}95 of the vehicle location error in Fig.~\ref{fig:rmse_comparison_town02}, they further confirm the superiority of the proposed cooperative localization framework. 
Indeed, the location error of \ac{cls-mpnn} that integrates the detections provided by PointPillars and Part-$\text{A}^2$ is lower than 0.99 m and 1.17 m in 95\% of the cases, respectively, while GNSS-KF and C-SLAM attain 6.36 m and 1.36 m, respectively.
\textcolor{black}{Therefore, \ac{cls-mpnn} enhances the localization accuracy in the considered scenario compared to GNSS-KF and C-SLAM.}
Besides obtaining more accurate position estimates, \ac{cls-mpnn} is more communication-efficient compared with C-SLAM as only the estimated positions of vehicles and objects bounding boxes are exchanged with the \ac{mec} while C-SLAM requires vehicles to share also the raw LiDAR point cloud. 
Focusing on the comparison between \ac{o-cls-mpnn} and \ac{cls-mpnn}, some performance loss is experienced. 
This should not be surprising as \ac{o-cls-mpnn} detects all possible poles recognizable by the LiDAR sensors while also carrying out perfect \ac{da}. 
Nevertheless, the gap between \ac{cls-mpnn} implemented using the detections provided by PointPillars and \ac{o-cls-mpnn} is only $6$ cm.

\begin{figure}
    \centering
    \includegraphics[width=0.95\linewidth]{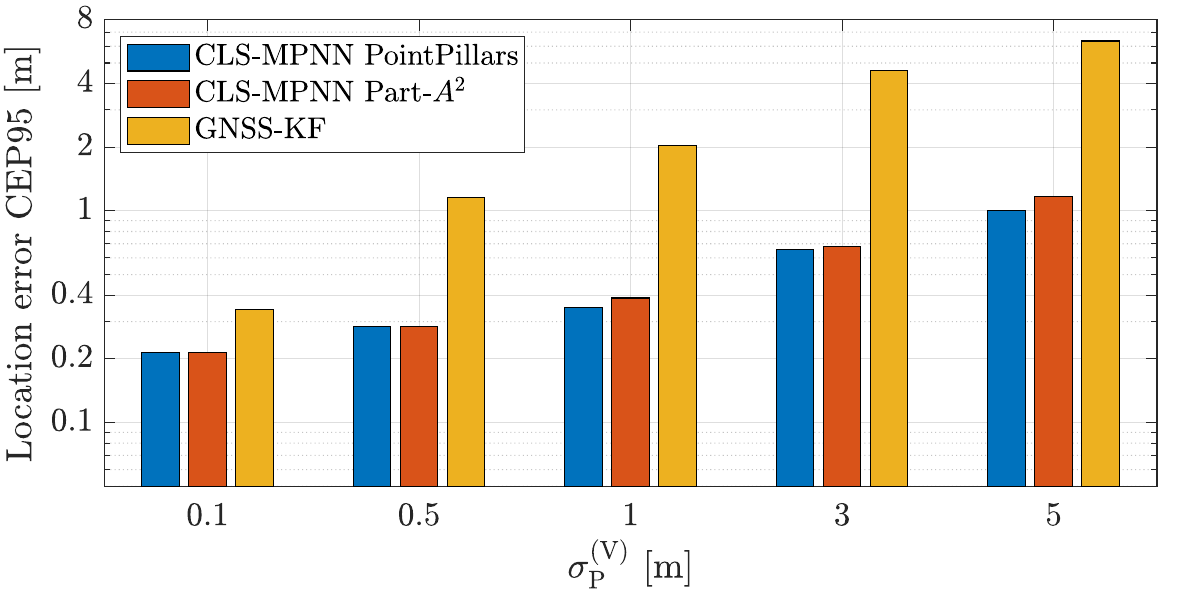}
    \vspace{-3mm}
    \caption{\textcolor{black}{CEP95 values of CLS-MPNN and GNSS-KF vs GNSS accuracy.}}
    \label{fig:gnss_accuracy_comparison}
\end{figure}
\begin{figure}
\vspace{-3mm}
    \centering
    \includegraphics[width=0.95\linewidth]{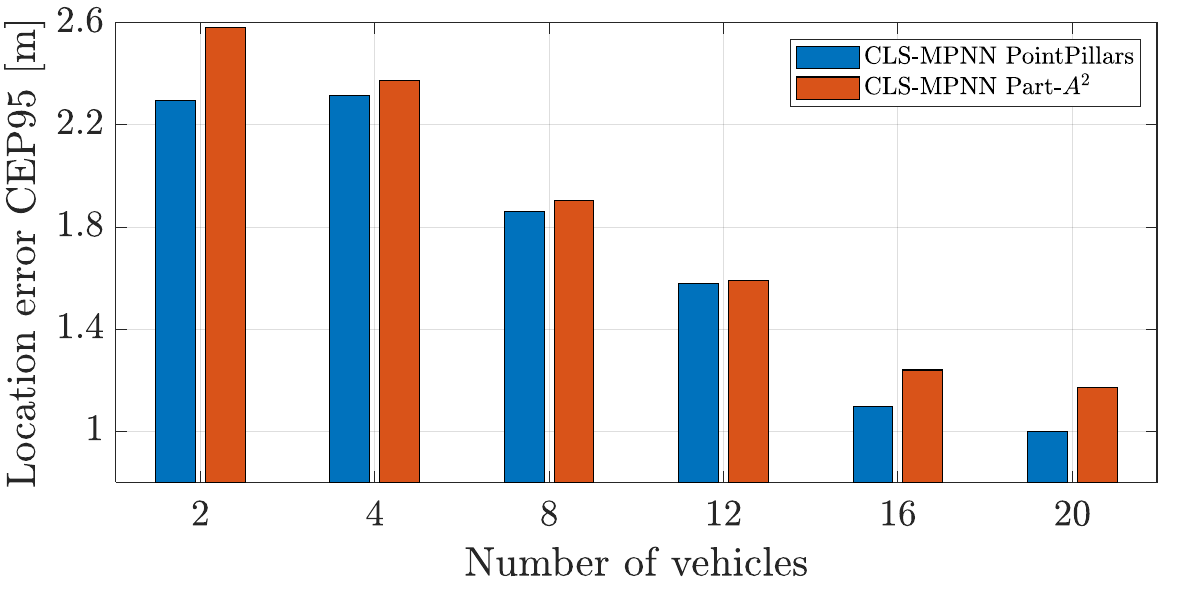}
    \vspace{-3mm}
    \caption{\textcolor{black}{CEP95 values of the vehicle location error vs the number of vehicles.}}
    \label{fig:results_cavs}
    \vspace{-4mm}
\end{figure}

\subsubsection{\textcolor{black}{Impact of GNSS accuracy}}
\textcolor{black}{To complement the analysis shown before,  we here evaluate the impact of varying the \ac{gnss} positioning uncertainty $\sigma_{\text{P}}^{(\text{V})}$.
Fig. \ref{fig:gnss_accuracy_comparison} reports the vehicle location error CEP95 attained by \ac{cls-mpnn} and GNSS-KF when $\sigma_{\text{P}}^{(\text{V})}$ ranges from 0.1 m up to 5 m. 
The results show that \ac{cls-mpnn} is particularly useful for enhancing the positioning performances when the GNSS is highly inaccurate.
Indeed, when $\sigma_{\text{P}}^{(\text{V})} < 1$ m the average improvement brought by \ac{cls-mpnn} with respect to GNSS-KF is 57\%, while for $\sigma_{\text{P}}^{(\text{V})} \geq 1$ m the same improvement is roughly 87\%.
Besides, the analysis highlights that \ac{cls-mpnn} is advantageous even when the GNSS accuracy is high. }

\subsubsection{\textcolor{black}{Impact of number of vehicles}}
\textcolor{black}{In Fig.~\ref{fig:results_cavs}, we analyze how the \ac{cep}95 of the vehicle location error is influenced by the number of vehicles in the scenario.
We compare the results for the \ac{cls-mpnn} using PointPillars and Part-$\text{A}^2$ when $N_{v}$ ranges from $2$ up to $20$.
The achieved results show a diminishing return in terms of positioning accuracy as the number of connected vehicles decreases: the localization error is more than doubled when passing from $N_{v}= 20$ to $N_{v}= 2$.
Nevertheless, even when $N_{v} = 2$, the proposed approach is still able to largely outperform GNSS-KF regardless of the detector employed.}

\begin{figure}
    \centering
    \includegraphics[width=0.95\linewidth]{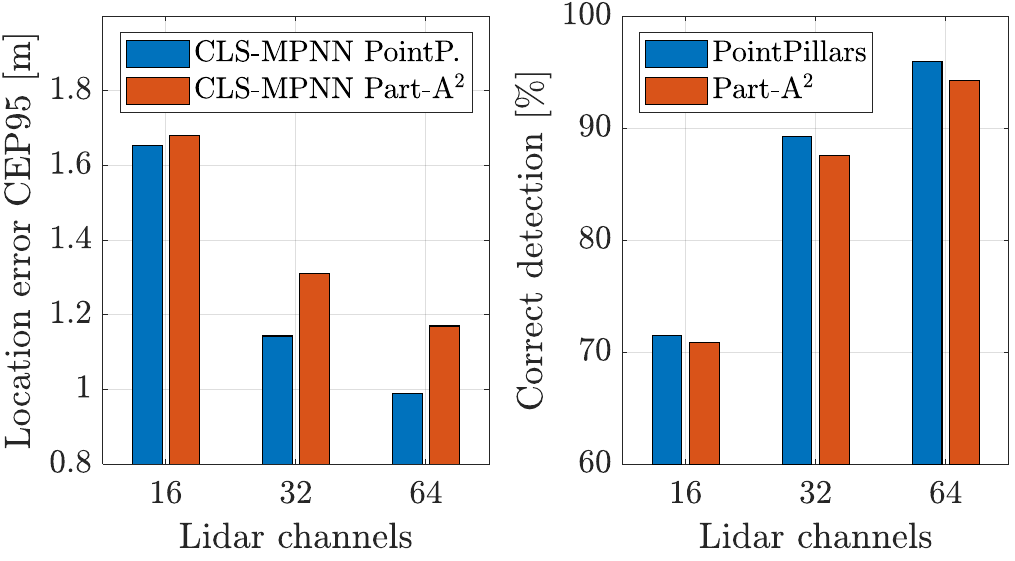}
    \vspace{-3mm}
    \caption{\textcolor{black}{CEP95 values of the vehicle location error and percentage of correct detection  vs  LiDAR resolution.} }
    \label{fig:lidar_quality}
    \vspace{-3mm}
\end{figure}

\subsubsection{\textcolor{black}{Impact of LiDAR resolution}}
\textcolor{black}{We  here analyze how the resolution of the LiDAR (i.e., number of channels) affects the localization performances of \ac{cls-mpnn}. 
In the left panel of Fig.~\ref{fig:lidar_quality} we report the \ac{cep}95 values of vehicle location accuracy when using LiDARs with $16$, $32$, and $64$ channels, while the right panel highlights the percentage of correct detections obtained by PointPillars and Part-$\text{A}^2$.
\textcolor{black}{Note that the 3D object detectors are not retrained to account for the lower LiDAR resolution.}
We observe that as the number of channels decreases, the positioning accuracy decreases as well, regardless of the detector employed.
This should be expected since a lower resolution makes the detection process more difficult, leading to fewer objects detected and a consequent increase of the \ac{cep}95 value.
To improve the object detection performances, super-resolution methods could be adopted~\cite{lidar_super_res}.
We remark that \ac{cls-mpnn} still outperforms GNSS-KF even when using low-resolution LiDAR sensors.}

\begin{table}[!t]
\renewcommand{\arraystretch}{1.3}
	\setlength{\tabcolsep}{4pt}
    \caption{\textcolor{black}{Analysis on the impact of false positives }}
     \vspace{-2mm}
    \label{tab:fp_minimization}
    \centering
    \begin{tabular}{|c|c|c|c|c|}
    \hline
    \multirow{3}{*}{\textcolor{black}{\textbf{Detector}}} & \textcolor{black}{\textbf{Original}}  & \textcolor{black}{\textbf{Strategy}}  & \textcolor{black}{\textbf{Strategy}} & \textcolor{black}{\textbf{Strategy}} \\
    & \textcolor{black}{\textbf{false positives}} & \textcolor{black}{\textbf{1}} & \textcolor{black}{\textbf{2}} & \textcolor{black}{\textbf{3}} \\
    & \textcolor{black}{[\%]} & \textcolor{black}{[\%]} & \textcolor{black}{[\%]} & \textcolor{black}{[\%]} \\ \hline
    \textcolor{black}{PointPillars} & \textcolor{black}{12.59} & \textcolor{black}{0.88} & \textcolor{black}{0.48} & \textcolor{black}{0.18} \\  
    \textcolor{black}{Part-$\text{A}^2$}  & \textcolor{black}{16.72} & \textcolor{black}{3.13} & \textcolor{black}{1.75} & \textcolor{black}{0.38} \\ \hline
    \end{tabular}
 \vspace{-2mm}
\end{table}

\subsubsection{\textcolor{black}{Impact of false positives}} 
\begin{figure}
    \centering
    \includegraphics[width=0.95\linewidth]{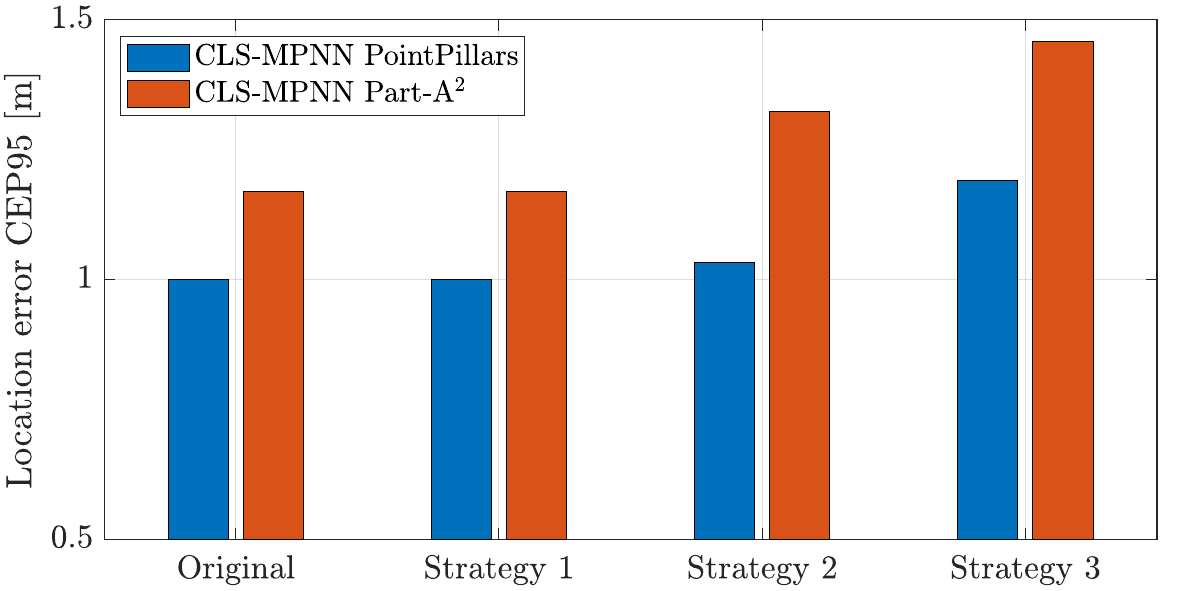}
    \vspace{-3mm}
    \caption{\textcolor{black}{CEP95 values attained by CLS-MPNN after applying the filtering strategies of FPs.}}
    \label{fig:fp_analysis}
\end{figure}

\textcolor{black}{Even though the proposed method is not formulated to deal with \acp{fp}, we here highlight how the detections can be filtered to minimize the false alarms without heavily impacting the performances of \ac{cls-mpnn}.
To reduce the number of \acp{fp}, we employ three filtering strategies.
The first one (Strategy 1) retains only bounding boxes detected by multiple vehicles at the same time instant, whereas the other two take as input the output of the first one and remove boxes having a score probability less than 0.15 (Strategy 2) and 0.25 (Strategy 3). 
Table~\ref{tab:fp_minimization} and Fig. \ref{fig:fp_analysis} report the percentage of \acp{fp} and the vehicle location error CEP95 of \ac{cls-mpnn} before and after applying the three filtering strategies for both PointPillars and Part-$\text{A}^2$, respectively. 
Analyzing the results, the first strategy heavily limits the number of \acp{fp} without reducing the final accuracy.
On the other hand, the other two policies introduce marginal accuracy drops but allow to almost completely cancel out the false alarms, i.e., the \acp{fp} are less than 1\% of all the detections.
Indeed, removing bounding boxes that have a score lower than 0.25 reduces the percentage of \acp{fp} to $0.18$\% and to $0.38$\% for PointPillars and Part-$\text{A}^2$, respectively.
This in turn makes \ac{cls-mpnn} slightly less accurate, increasing its CEP95  to 1.19 m and 1.46 m for PointPillars and Part-$\text{A}^2$, respectively.}

\begin{figure}
\vspace{-3mm}
    \subfloat[\label{fig:times}]{
		\includegraphics[width=0.95\linewidth]{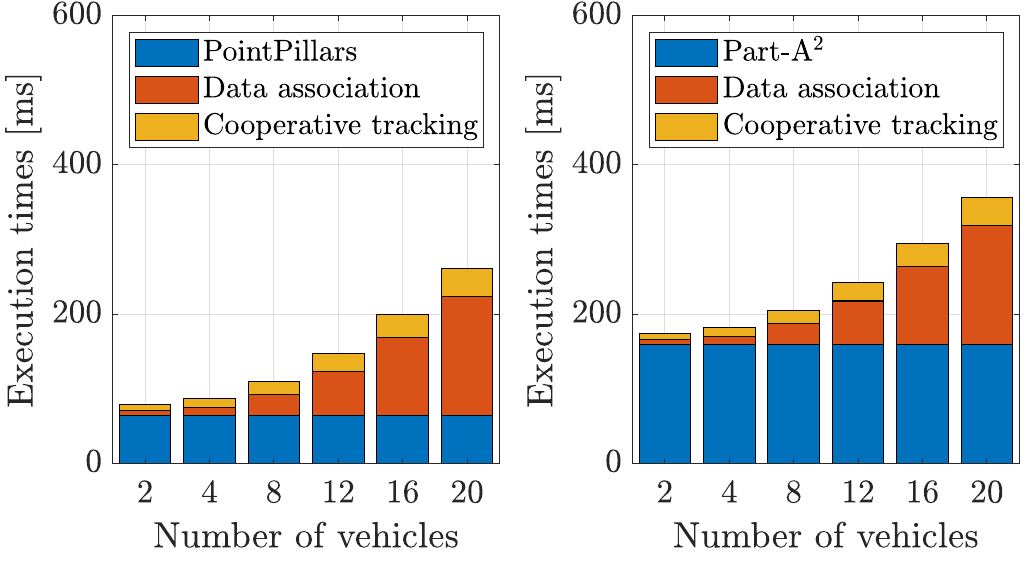}
	}

\vspace{-3mm}
\subfloat[\label{fig:memory}]{
		\includegraphics[width=0.95\linewidth]{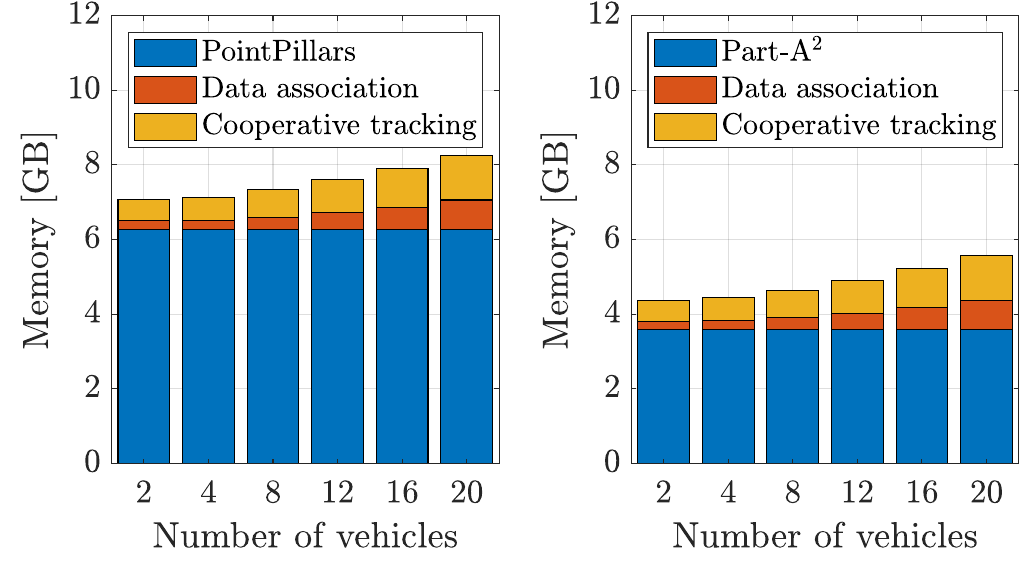}
	}
 \vspace{-1mm}
    \caption{\textcolor{black}{Execution time (a) and memory (b) required to run the proposed approach as a function of the number of vehicles.}}
    \label{fig:results_computations}
    \vspace{-3mm}
\end{figure}

\subsubsection{\textcolor{black}{Computational requirements}}
\textcolor{black}{We analyze the computational complexity for executing the proposed method by reporting in
Fig.~\ref{fig:times}  the computational time and in Fig.~\ref{fig:memory}  the associated memory utilization, with a breakdown on the individual contributions from the detector, \ac{da} algorithm and cooperative tracker. 
Results show that the most time-consuming stages for running the developed approach are the \ac{da} step and the 3D object detection process. For what concerns the memory utilization instead, a large memory portion should be devoted to storing the detectors. The figures also show that the times/memory required for running the \ac{da} and the cooperative tracking steps increase as the number of connected vehicles grows. This is reasonable since, as opposed to the object detection process which is performed onboard the vehicles, a larger number of vehicles leads to a more complex cooperation among vehicles.} 

\begin{figure}
    \centering
    \includegraphics[width=0.95\linewidth]{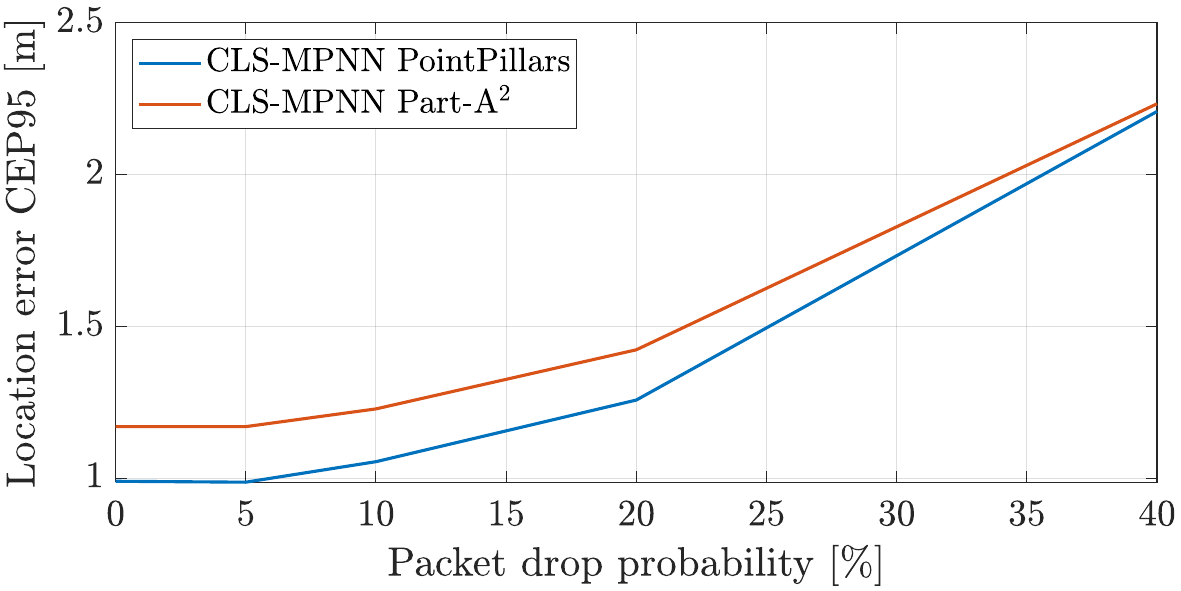}
    \vspace{-4mm}
    \caption{\textcolor{black}{Impact of packet drops on the vehicle location error CEP95.}}
    \label{fig:results_drop}
    \vspace{-4mm}
\end{figure}

\subsubsection{\textcolor{black}{Impact of network reliability}}
\textcolor{black}{We here evaluate the robustness of \ac{cls-mpnn} against unreliable networks by introducing packet drop events in the cooperative framework. 
They are assumed to be independent across vehicles and we consider that if a packet drop occurs between the $v$-th vehicle and the \ac{mec} all the transmitted information (i.e., the position of the $v$-th vehicle and the poles bounding boxes detect by it) is lost.
\textcolor{black}{Herein, we do not rely on \ac{v2x} networking simulators
as we want to benchmark \ac{cls-mpnn} under extreme conditions.
Such conditions are likely not to be encountered when utilizing \ac{v2x} simulation frameworks but are useful to highlight the potential limits of the proposed approach.}
Fig.~\ref{fig:results_drop} shows the location error \ac{cep}95 achieved by \ac{cls-mpnn} when the packet drop probability ranges from 0\% (i.e., no packet drops) to 40\%.
The results show that the \ac{cls-mpnn} is negligibly affected if the packet drop probability is lower than 10\% regardless of the detector considered. For larger packet drop occurrences, a strong reduction on the accuracy is experienced. }

\vspace{-4mm}
\subsubsection{\textcolor{black}{Impact of LiDAR sensing range}}
\begin{figure}[!t]
	\centering
	\includegraphics[width=0.95\linewidth]{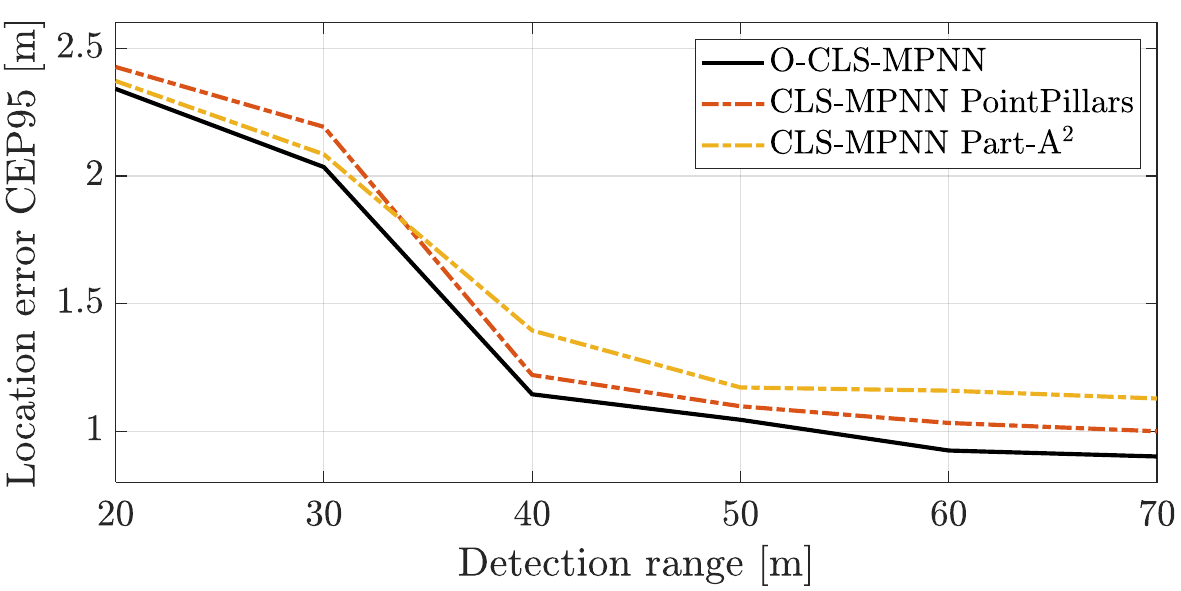}
	\vspace{-4mm}
	\caption{CEP95  values of vehicle location error in \textit{Town02} for \ac{o-cls-mpnn} and \ac{cls-mpnn} with PointPillars and Part-$\text{A}^2$ for different LiDAR sensing ranges. }
 \vspace{-4mm}
	\label{fig:rmse_ranges_town02}
\end{figure}

Fig.~\ref{fig:rmse_ranges_town02} compares the performances on vehicle positioning obtained by \ac{cls-mpnn} integrating the PointPillars and Part-$\text{A}^2$ detections with the ones achieved by \ac{o-cls-mpnn} considering different LiDAR sensing ranges $R_s$ ranging from $20$~m up to $70$~m, i.e., the case considered previously. 
Comparing the results, PointPillars achieves a lower \ac{cep}95 compared with Part-$\text{A}^2$ in nearly all cases.
This should be expected since, in this scenario, PointPillars detects more poles compared to Part-$\text{A}^2$ (as highlighted in Table \ref{tab:detection_town02}).
Comparing the results for different values of $R_s$, the performances of all methods do not deteriorate too much for $R_s \geq 40$~m, while for $R_s < 40$~m a large performance drop occours.
Therefore, the proposed approach could be employed with lower-grade LiDAR sensors provided that $R_s \geq 40$~m. 

\begin{figure}[!t]
    \centering
    \includegraphics[width=0.95\linewidth]{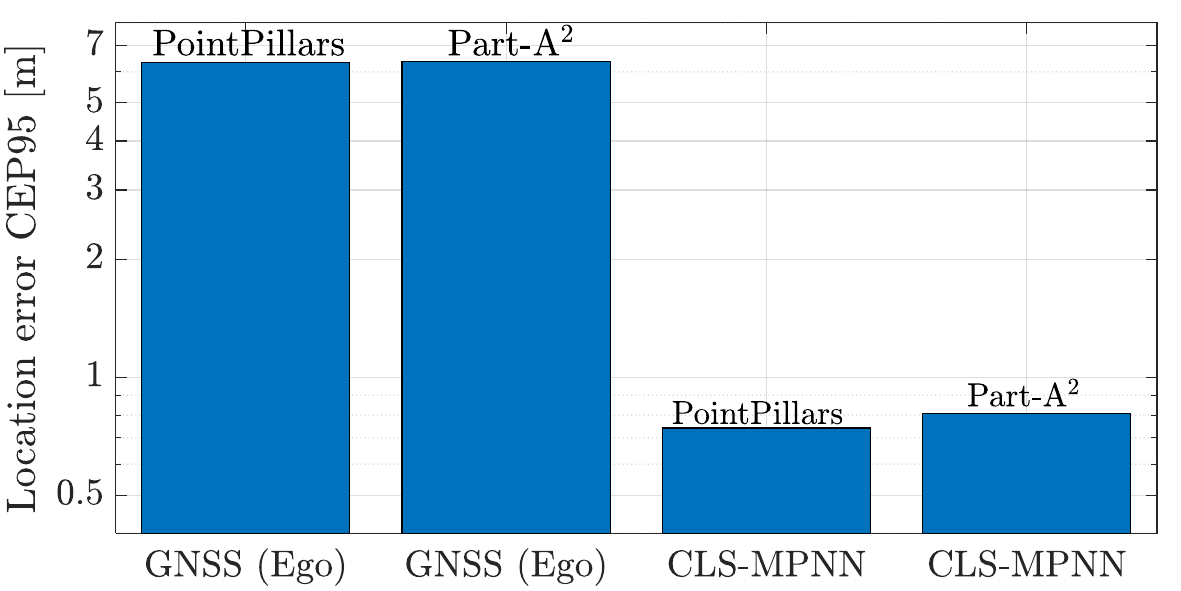}
    \vspace{-3mm}
    \caption{\textcolor{black}{CEP95 of pole location error for different localization methodologies.}}
    \label{fig:results_poles}
    \vspace{-4mm}
\end{figure}

\subsubsection{\textcolor{black}{Object position accuracy}}
\textcolor{black}{As a final analysis, we evaluate the  positioning accuracy of poles, comparing the proposed approach  with an ego-sensing solution.
In the latter, vehicles obtain their positions via the GNSS-KF approach and use this information jointly with the LiDAR detections to localize the poles in a global reference frame.
The results of this comparison are highlighted in Fig.~\ref{fig:results_poles} considering the detections provided by PointPillars and Part-$\text{A}^2$ detectors.
The \ac{cep}95 location error of \ac{cls-mpnn} is largely reduced compared to the one attained by the ego sensing solution for both detectors. This highlights the benefits of employing \ac{cls-mpnn} to also improve the localization performances of the poles with respect to non-cooperative sensing schemes.}

\subsection{Town10 results}
\label{subsec:town10_results}

In this section, we evaluate \ac{cls-mpnn} in a new scenario where the \acp{nn} have not been trained on. 
This new evaluation is instrumental for characterizing the developed method under unseen conditions.
For this scope, we employ the \textit{Town10} map of CARLA, a more complex environment comprising an overpass and roads with multiple lanes, where 20 LiDAR-equipped vehicles move in the environment.
The same calibration and LiDAR parameters employed in Sec. \ref{subsec:town02_results} and in Sec. \ref{sec:nn_results} are also used here. 
The simulation lasts $300$ seconds and comprises $1500$ snapshots each sampled at $T_s = 0.2$ s. 
As done before, we report in Table \ref{tab:detection_town10} the percentages of correct detections, missed detections, and \acp{fp} attained by PointPillars and Part-$\text{A}^2$.
\textcolor{black}{For this map, Part-$\text{A}^2$ is observed to generalize better as it provides fewer missed detections compared to PointPillars, despite detecting a larger number of false positives.
This is in line with the results provided in~\cite{double_stage_generalization} as they show that double-stage detectors (e.g., Part-$\text{A}^2$) outperform single-stage ones (i.e., PointPillars) when the testing conditions change. 
Compared with the values provided in Table \ref{tab:detection_town02}, \acp{fp} and missed detections are higher, indicating that fine-tuning the detectors on this new map may help improve the performances.}


\begin{table}[!t]
	\renewcommand{\arraystretch}{1.3}
	\setlength{\tabcolsep}{4pt}
	\caption{Detection Performances Analysis in Town10}
	\label{tab:detection_town10}
 \vspace{-2mm}
	\centering
	\begin{tabular}{|c|c|c|c|}
		\hline
		\multirow{2}{*}{\textbf{Method}}  & \textbf{Correct detections} & \textbf{Missed detections} & \textbf{False positives} \\
		& [\%] & [\%] & [\%] \\ \hline
		PointPillars & 83.59 & 16.40 & 9.21 \\
		Part-$\text{A}^2$ & 87.89 & 12.11 & 14.26 \\ \hline
	\end{tabular}
 \vspace{-4mm}
\end{table}

\begin{figure*}[tb]
	\centering
	\subfloat[\label{fig:rmse_map_gps_town10}]{
		\includegraphics[width=0.32\linewidth]{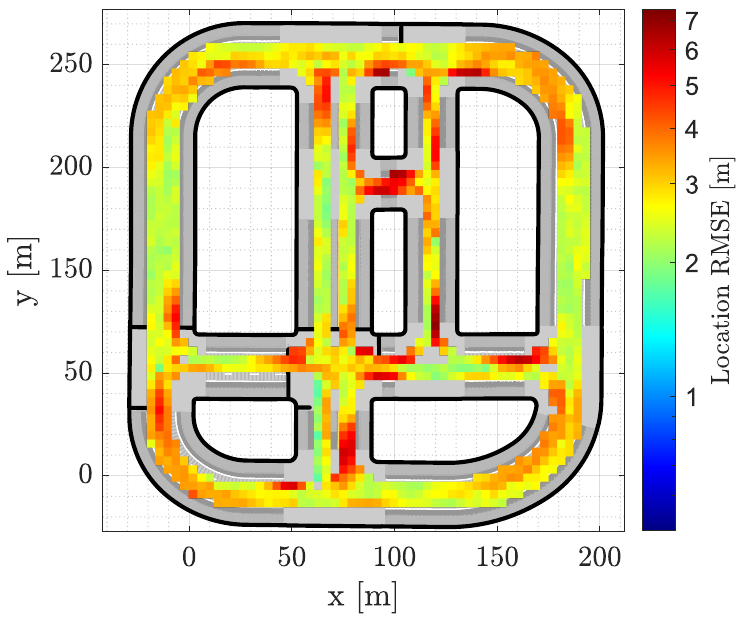}
	}
	\subfloat[\label{fig:rmse_map_oracle_town10}]{
		\includegraphics[width=0.32\linewidth]{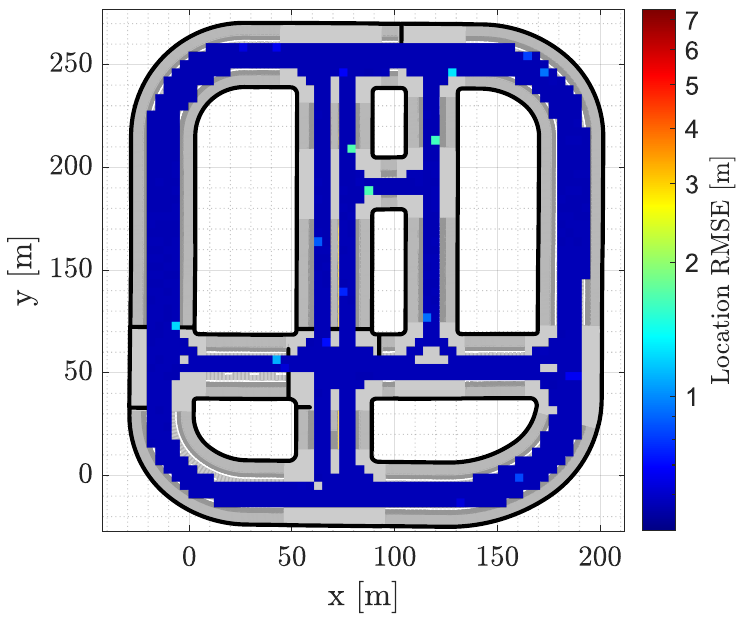}
	}
	\subfloat[\label{fig:rmse_map_pointpillars_town10}]{
		\includegraphics[width=0.32\linewidth]{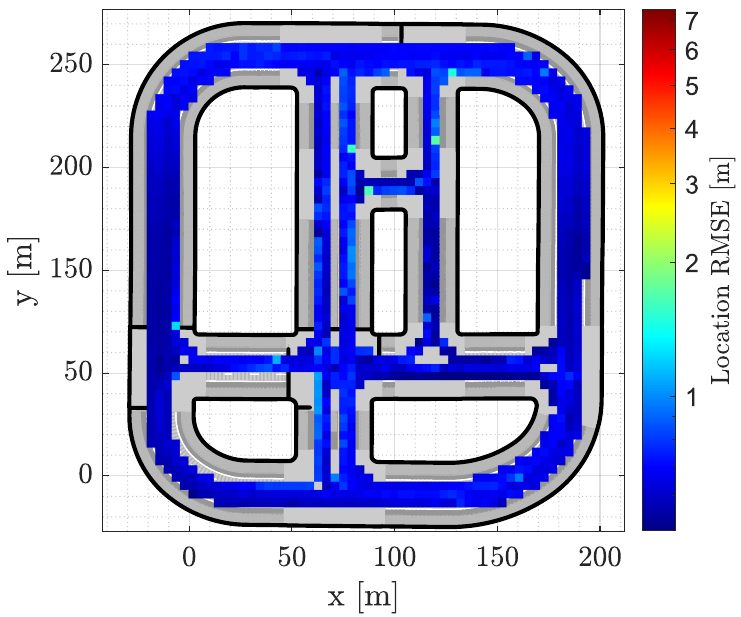}
	}
	
	\vspace{-3mm}
	\subfloat[\label{fig:rmse_map_parta2_town10}]{
		\includegraphics[width=0.32\linewidth]{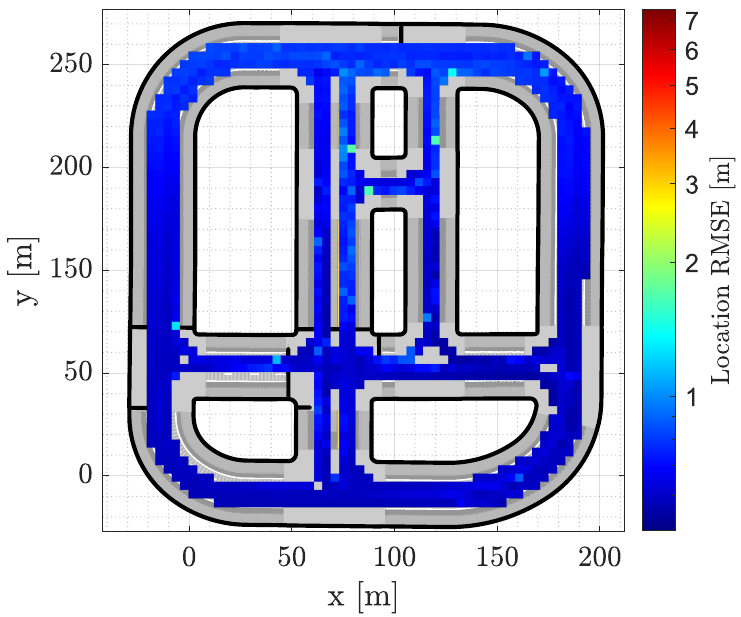}
	}
	\subfloat[\label{fig:rmse_comparison_town10}]{
	\hfill
	\includegraphics[width=0.355\linewidth]{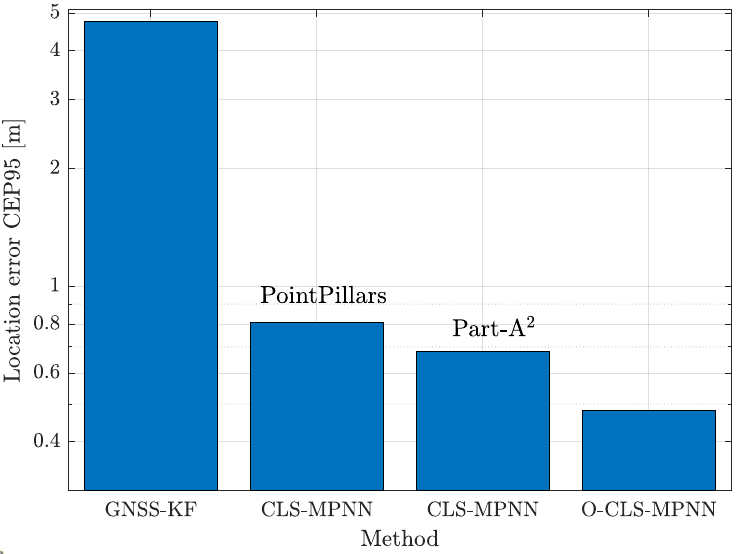}
}
	\vspace{-1mm}
	\caption{\textcolor{black}{Spatial RMSE of vehicle positions over the \textit{Town02} simulation map:  (a) GNSS-KF. (b) \ac{o-cls-mpnn} (perfect \ac{da} and LiDAR sensing) (c) \ac{cls-mpnn} with PointPillars, (d) \ac{cls-mpnn} with Part-$\text{A}^2$, (e) comparison of the CEP95 of the location error for all methods.}}
	\label{fig:results_comparison_town10}
	\vspace{-6mm}
\end{figure*}
\begin{figure}[!t]
	\centering
	\includegraphics[width=\linewidth]{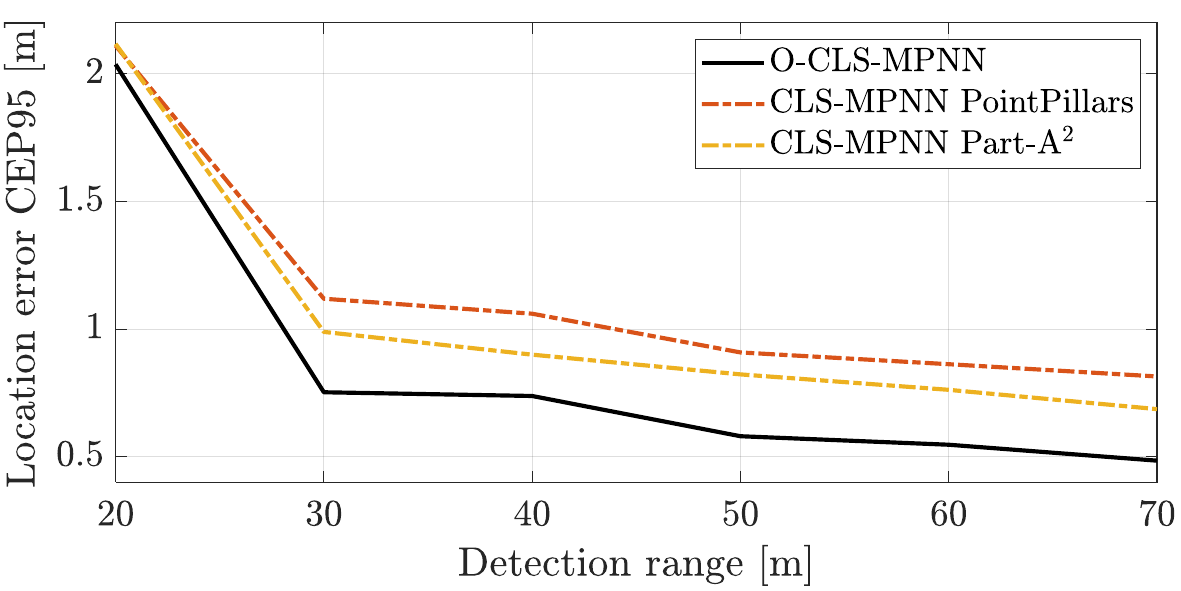}
	\vspace{-7mm}
	\caption{Vehicle location error CEP95 in \textit{Town10} for \ac{o-cls-mpnn} and \ac{cls-mpnn} with PointPillars and Part-$\text{A}^2$ for different LiDAR sensing ranges. }
	\label{fig:rmse_ranges_town10}
 \vspace{-5mm}
\end{figure}


Fig.~\ref{fig:results_comparison_town10} shows the spatial \ac{rmse} computed all over the map for GNSS-KF (Fig.~\ref{fig:rmse_map_gps_town10}), \ac{o-cls-mpnn} (Fig.~\ref{fig:rmse_map_oracle_town10}), the proposed \ac{cls-mpnn} method considering the detections provided by PointPillars (Fig.~\ref{fig:rmse_map_pointpillars_town10}) and by Part-$\text{A}^2$ (Fig.~\ref{fig:rmse_map_parta2_town10}), while Fig.~\ref{fig:rmse_comparison_town10} shows the CEP95 of the location error obtained by all methods.
The results of this analysis are in line with the findings of the previous section: \ac{cls-mpnn}  largely outperforms GNSS-KF regardless of the detector considered while it approaches the performances of \ac{o-cls-mpnn}.
\textcolor{black}{Again, some spots can be noticed where the improvement brought by \ac{cls-mpnn} is only marginal due to a low cooperation level between vehicles.}
With respect to the previous case, the CEP95 of the location error is generally lower for all methods.
This is related to the fact that a larger number of poles is present on this map with respect to the ones present in \textit{Town02} that can be exploited for cooperative vehicle positioning refinement.
Another difference with respect to the previous results is that the performance gap between \ac{cls-mpnn} and \ac{o-cls-mpnn} is higher, possibly due to the larger number of missed detections. 
Nevertheless, \ac{cls-mpnn} is shown to generalize to this new scenario as it provides accurate vehicle positioning.
Contrarily to before, now the detections provided by Part-$\text{A}^2$ should be chosen for maximizing the vehicle position accuracy. 

To conclude the analysis, we report in Fig.~\ref{fig:rmse_ranges_town10} the CEP95 of the location error obtained by \ac{cls-mpnn} implemented with the detections provided by PointPillars and Part-$\text{A}^2$ and the one achieved by \ac{o-cls-mpnn} for varying LiDAR sensing ranges. 
Under this map, the predictions provided by Part-$\text{A}^2$ lead to a lower positioning error compared to the ones provided by PointPillars for nearly all sensing ranges. 
Therefore, despite Part-$\text{A}^2$ providing less accurate performances in \textit{Town02}, it should be preferred compred to PointPillars as it generalizes to unseen scenarios and attain higher localization accuracy when integrated with \ac{cls-mpnn}. 
Notwithstanding this aspect, it is advisable to partially retrain or fine-tune the models to enhance even more the positioning accuracy as the performance gap between \ac{o-cls-mpnn} and \ac{cls-mpnn} is higher when compared to the same gap shown in Fig.~\ref{fig:rmse_ranges_town02}.

\subsection{\textcolor{black}{Considerations on the communication overhead}}
\label{subsec:communication_overhead}

\textcolor{black}{We here compare the communication overhead of the proposed \ac{cls-mpnn} with the one required by C-SLAM.
An upper bound on the communication overhead (in bits) required by each vehicle to run \ac{cls-mpnn} is $N_{vd} E_{v} + N_{od} N_{c} N_{o} E_{o}$.
The first term encodes the number of bits used for exchanging the 3D vehicle position, where $N_{vd} = 3$ and $E_{v}$ denotes the number of bits encoding the position, while the second refers to the exchange of the detected objects, where $N_{c} = 8$ is the number of 3D corners of the bounding box, $N_{od} = 3$, and $E_{o}$ is the number of bits used for encoding the information.
A typical value for $E_{v}$ and $E_{o}$ is 32 bits.
On the other hand, the communication overhead of C-SLAM is upper bounded by $N_{vd} E_{v}+N_{pd} N_{\ell p} E_{\ell p}$. The first term is the same as before, while the second highlights the bits required for sharing the raw point cloud, with $N_{pd} =3$ and $E_{p}$ denotes the number of bits for representing a single point of the cloud.
Clearly, \ac{cls-mpnn} is advantageous over C-SLAM when $N_{pd} N_{\ell p} E_{\ell p} \gg N_{od} N_{c} N_{o} E_{o}$.
This is typically verified considering LiDAR sensors with at least 16 channels.
Indeed, on average, such sensors output $N_{\ell p} = 300000$ points for each frame~\cite{velodyne}, thus, the term $N_{pd} N_{\ell p} E_{\ell p}$ corresponds to 0.28 Mbits (assuming that point clouds are compressed using G-PCC). 
To get the same communication overhead for \ac{cls-mpnn}, $N_{o}$ should be at least 375, which is far beyond the maximum number of output objects from 3D detectors (limited to 50). 
In this condition, the communication overhead of CLS-MPNN is reduced by 7.5 times compared to C-SLAM.}

\section{Conclusion}
\label{sec:conclusions}

In this paper, we proposed a novel data-driven cooperative localization and sensing framework, referred to as \ac{cls-mpnn}, for \ac{gnss} augmentation in complex urban environments. 
The framework leverages a \ac{nn}, i.e., a 3D object detector, at the vehicles for recognizing and localizing static objects in the surroundings. 
The detections are then aggregated at a centralized unit, namely a \ac{mec}, that performs a \ac{mpnn}-based data association procedure to coherently combine the measurements provided by multiple vehicles and referring to a same object.
Once the data association is solved, the \ac{mec} employs a cooperative Bayesian tracking algorithm to improve the objects' localization accuracy and implicitly refines the vehicle positioning as well. 

We employed two synthetically-generated, yet realistic, urban scenarios to assess the proposed cooperative method.
The first one is used to assess the performances of the employed single-stage/double-stage object detectors, the \ac{mpnn} data association accuracy, and, finally, to evaluate the performances of the overall cooperative localization framework. 
The second one, instead, is utilized to characterize the ability of the developed technique to generalize to new unseen conditions.
The results show that \ac{cls-mpnn} is able to substantially outperform a conventional non-cooperative solution based on stand-alone \ac{gnss} and a state-of-the-art cooperative \ac{slam} approach. 
Furthermore, when the scenario is modified with respect to the one used during training, a marginal performance loss is experienced by \ac{cls-mpnn} due to more missed detections. 
Nevertheless, \ac{cls-mpnn} is still able to substantially reduce the positioning errors with respect to the ego positioning solution while approaching the performances of an oracle system that perfectly associates and detects all objects present in the surroundings of the vehicles.
The analysis also shows that double-stage 3D object detectors are to be preferred compared to single-stage ones as they show better generalization performances to new scenarios.

\textcolor{black}{Future research activities will target the assessment  in real-world driving scenarios as well as the development of end-to-end cooperative learning strategies where 3D object detection and data association tasks are carried out by a single \ac{nn} possibly designed to also support fully distributed implementations.
Besides, the \ac{mpnn}-based data association strategy could be extended to deal with the false positives.} 

\bibliographystyle{IEEEtran}
\bibliography{IEEEabrv,bib_formatted}

\vfill

\end{document}